\documentclass[a4paper,fleqn,usenatbib]{mnras}

\usepackage[T1]{fontenc}
\usepackage{ae,aecompl}

\usepackage{graphicx}	
\usepackage{amsmath}	
\usepackage{amssymb}	
\usepackage{hyperref}

\def\VEV#1{\left\langle #1\right\rangle} 
  
\title[High-z Galaxies and Feedback]{A Minimalist Feedback-Regulated Model for Galaxy Formation During the Epoch of Reionization}

\author[Furlanetto et al.]{
Steven R. Furlanetto,$^{1}$\thanks{E-mail: sfurlane@astro.ucla.edu (SRF)},
Jordan Mirocha,$^{1}$ 
Richard Mebane,$^{1}$ 
\& Guochao Sun$^{2}$\\
$^{1}$Department of Physics \& Astronomy, University of California, Los Angeles, Los Angeles, CA 90095, USA\\
$^{2}$Cahill Center for Astronomy and Astrophysics, California Institute of Technology, Pasadena, CA 91125, USA
 \\
}

\date{Accepted XXX. Received YYY; in original form ZZZ}

\pubyear{2015}

\begin{document}
\label{firstpage}
\pagerange{\pageref{firstpage}--\pageref{lastpage}}
\maketitle

\begin{abstract}
Near-infrared surveys have now determined the luminosity functions of galaxies at $6 \la z \la 9$ to impressive precision and identified a number of candidates at even earlier times. Here we develop a simple analytic model to describe these populations that allows physically-motivated extrapolation to earlier times and fainter luminosities.  We assume that galaxies grow through accretion onto dark matter halos, which we model by matching halos at fixed number density across redshift, and that stellar feedback limits the star formation rate.  We allow for a variety of feedback mechanisms, including regulation through supernova energy and momentum from radiation pressure. We show that reasonable choices for the feedback parameters can fit the available galaxy data, which in turn substantially limits the range of plausible extrapolations of the luminosity function to earlier times and fainter luminosities: for example, the global star formation rate declines rapidly at $z \ga 10$, but the bright galaxies accessible to observations decline much faster than the total. Deviations from our predictions would provide evidence for new astrophysics within the first generations of galaxies. We also provide predictions for galaxy measurements by future facilities, including JWST and WFIRST.
\end{abstract}




\section{Introduction} \label{intro}

One of the major goals of extragalactic astrophysics is to map the formation of the first galaxies and their evolution into the mature objects we observe at $z \la 1$.  Much of that story has been revealed over the last two decades, with observational astronomers pushing the frontiers past the peak era of star formation, where we are now developing a sophisticated understanding of galaxy evolution. Over the past several years, the Wide Field Camera 3 on the \emph{Hubble Space Telescope} (HST) has enabled a census of star formation in the first billion years of the Universe's history at $z \sim 6$--8 \citep{mclure13,schenker13,finkelstein15,bouwens15,atek15,bowler16}, with a few sources now detected at $z \sim 9$--11 \citep{oesch13,oesch15,bouwens15b, ishigaki15,mcleod15,mcleod16}.

While we know little more than the abundance of bright galaxies at this time, enough progress has been made to begin modeling the contents, formation, and evolution of these early generations of galaxies. There is (as yet) no evidence that these galaxies have particularly unusual stars \citep{dunlop12, bouwens14}, though there are hints that the detailed processes of star formation differ \citep{munoz14-cii, capak15}.  Qualitatively, their behavior is consistent with galaxies at lower redshifts \citep{behroozi13, sun16}, with a peak in the star formation efficiency at a halo mass $m_h \sim 10^{11} \ M_\odot$ and $\la 10\%$ of the baryons converted into stars. 

With this basic picture in hand, extensive theoretical work is underway to understand the processes driving these sources. Much of this has been performed in the context of detailed numerical simulations (e.g., \citealt{jaacks12,trac15,feng16,gnedin16,waters16}) and semi-analytic models (e.g., \citealt{dayal14, liu16, mutch16,yue16}). Here we take a different approach, developing a flexible, analytic model of galaxy formation from a set of simple, transparent physical assumptions. While such a model cannot be used to describe the detailed properties of the galaxies, it provides a basis for understanding the qualitative features of the galaxy populations (which are, so far, the limits of our observations). Importantly, it also allows \emph{physically-motivated} extrapolation to both early times and faint luminosities, guiding our understanding of empirical extrapolations that are now widely used \citep{robertson13, robertson15, mason15, visbal15, mashian16, sun16, mirocha16}.  Such extrapolations are crucial not only to observations of galaxies at earlier times (with, for example, the future \emph{James Webb Space Telescope}, or JWST, and the \emph{Wide-Field Infrared Survey Telescope}, or WFIRST) but also to measurements of the reionization process, which depends on the \emph{cumulative} number of photons emitted by the entire galaxy population.

The model presented below is similar to those of \citet{dayal14} and \citet{trenti10}, though simpler and more flexible than both, and is inspired by the basic physics driving galaxy formation models at lower redshifts.  We assume galaxies grow inside dark matter halos and form stars as those halos accrete more material. We set the star formation rate by assuming that stellar feedback, through radiation or supernovae, ejects the remaining inflowing material. We will show that such a framework can fit all available observations with reasonable parameter choices and that calibrating to those observations substantially limits the expected behavior at faint luminosities and earlier times.  Deviations from these extrapolations are therefore likely to provide indications of new astrophysics -- new stellar populations or star formation channels qualitatively different from those that dominate in the later universe.

This paper is organized as follows. In Section \ref{dmhalo}, we describe our treatment of dark matter halos, including their abundance and growth, which forms the basis for our calculations. In Section \ref{sf-model}, we describe the feedback model that specifies the star formation rates in our model. In Section \ref{sfrhist}, we compare our model to observations of distant galaxies and show predictions for the overall star formation rate.  In Section \ref{gals}, we derive some simple properties of the galaxy population in the context of our model. Finally, in Section \ref{model-implications} we consider some of the model's implications for reionization and metal production, and we conclude in Section \ref{disc}.

The numerical calculations here assume $\Omega_m = 0.308$, $\Omega_\Lambda = 0.692$, $\Omega_b = 0.0484$, $h=0.678$, $\sigma_8=0.815$, and $n_s=0.968$, consistent with the recent results of \citet{planck15-param}. Unless specified otherwise, all distances quoted herein are in comoving units.

\section{Dark Matter Halos} \label{dmhalo}

Our model for the galaxy population will use a few simple ingredients for the properties of galaxies, intentionally choosing the simplest options permitted by detailed observations and simulations. In this section, we will describe their cosmological context, the dark matter halos.

\subsection{The Halo Mass Function} \label{hmf}

We assume that galaxies inhabit dark matter halos, so the abundance of those halos is our first key ingredient.  We let $n_h(m,z)$ be the comoving number density of dark matter halos with masses in the range $(m, m+dm)$ at a redshift $z$.  Following convention, we write this as
\begin{equation}
n_h(m,z) = f(\sigma) { \frac{\bar{\rho}}{ m }} {d \ln \frac{(1/\sigma)}{dm}},
\label{eq:nhm}
\end{equation}
where $\bar{\rho}$ is the comoving matter density, $\sigma(m,z)$ is the rms fluctuation of the linear density field, smoothed on a scale $m$, and $f(\sigma)$ is a dimensionless function that parameterizes the barrier-crossing distribution of the linear density field. We assume for simplicity that this function is ``universal," though simulations show there may be small deviations \citep{tinker08-hmf}. For our default calculations, we take $f(\sigma)$ from a fit to recent high-$z$ cosmological simulations \citep{trac15}:
\begin{equation}
f(\sigma) = 0.150 \left[ 1 + \left( \frac{\sigma} {2.54} \right)^{-1.36} \right] e^{-1.14/\sigma^2}.
\label{eq:scorch-barrier}
\end{equation}
We have also recomputed our results for the \citet{sheth02} mass function, motivated by ellipsoidal collapse, in order to test their sensitivity to uncertainties in the halo mass function. In general terms, the two prescriptions are very similar at moderate masses over the range $6 \la z \la 10$, but equation~(\ref{eq:scorch-barrier}) has fewer high-mass halos than the older form. Differences in the normalization between model mass functions can simply be absorbed into the uncertain efficiency factors below.  Thus only the shape matters significantly, and such differences usually only occur on the massive end.

We will further assume that only halos above a specified minimum mass  $m_{\rm min}$ can form stars. There are several potential physical reasons for such a threshold. The first stars likely form through the cooling of molecular hydrogen, forming Pop III stars.  However, molecular hydrogen is destroyed by a weak UV background; the observed galaxy population is very unlikely to harbor such stars. In the spirit of extrapolating only the known galaxy population, we therefore ignore this possibility in the following. Thus, only if the halo's virial temperature $T_{\rm vir} \ga 10^4$~K can atomic line cooling in primordial gas allow efficient gas cooling and clumping to the densities required for star formation. Second, if the IGM has been photoheated through reionization, accretion is suppressed onto galaxies with $m \la 10^9 \ M_\odot$ \citep{noh14}. We will see below that this minimum mass is less important in a feedback-regulated model than one might naively expect from the steepness of the mass function, so we will generally use the atomic cooling threshold to compute $m_{\rm min}$. This corresponds to a mass $m \sim 10^8 \ M_\odot$.

\subsection{Accretion Rates} \label{acc}

The next ingredient of our model is how, on average, halos accrete matter, particularly in the high-$z$ regime.  We will assume that the rate at which galaxies form stars depends on this overall accretion rate, as described in the next section. For simplicity, we will ignore scatter in this relation: though mergers are an important part of halo growth, even at moderate redshifts the majority of matter is acquired through smooth ongoing accretion \citep{goerdt15}.

For a wide range of redshifts, and over moderately large halo masses, simulations have measured relations similar to \citep{mcbride09,goerdt15,trac15}
\begin{equation}
\dot{m} = A m^\mu (1+z)^\beta,
\label{eq:sim-gen}
\end{equation}
where $A$ is a normalization constant, $m$ is the halo mass, $\mu \ga 1$ in the simulations,\footnote{Specifically, for example, \citet{mcbride09} find $\mu=1.127$ at moderate redshifts, while \citet{trac15} find $\mu =1.06$ at $z \sim 6$--10 and $10^8 \la (m/M_\odot) \la 10^{13}$.} and $\beta$ converges to 5/2 at large redshifts. \citet{dekel13} (see also \citealt{neistein06, neistein08}) argued that this form can be generically understood through the extended Press-Schechter algorithm \citep{lacey93}: $\mu$ follows from the shape of the matter power spectrum, while $\beta \approx 5/2$ follows from the way the halo mass function depends on redshift (through the critical overdensity for collapse).

Unfortunately, the validity of this relation has not been tested at very high redshifts or at very small masses, ranges that are of interest to us. Some of our results are quite sensitive to such uncertainties, as we will be following halo growth over many e-foldings in mass. We therefore use a slightly more sophisticated model for our main results. We make the ansatz that halos maintain their overall number density as they evolve according to the mass function. The idea is similar to abundance matching \citep{vale04}, which populates the galaxy luminosity function with halos by matching number densities, and to studies that interpret the growth of the galaxy population by comparing  objects at fixed number density across many redshifts \citep{vandokkum10}. Our approach is the direct analog of the latter, as we match the halo mass function at different redshifts in order to determine the accretion rate. That is, we demand that at any given pair of redshifts $z_1$ and $z_2$ a halo has masses $m_1$ and $m_2$ such that
\begin{equation}
\int_{m_1}^\infty dm \, n_h(m|z_1) = \int_{m_2}^\infty dm \, n_h(m|z_2).
\label{eq:halo-mod-acc}
\end{equation}
We then assign an accretion rate $\dot{m}(m,z)$ by demanding that this is true for all redshifts and all halo masses.   We note that the analogous scatter-free mapping between galaxies across redshifts is too simple to explain detailed observations \citep{wellons16}, but given the limits of the current observations and the lack of better theoretical models at the masses and redshifts of interest, we choose the simplest possible approach that still guarantees self-consistency between the halo mass assembly histories and the mass function across many redshifts.

\begin{figure}
	\includegraphics[width=\columnwidth]{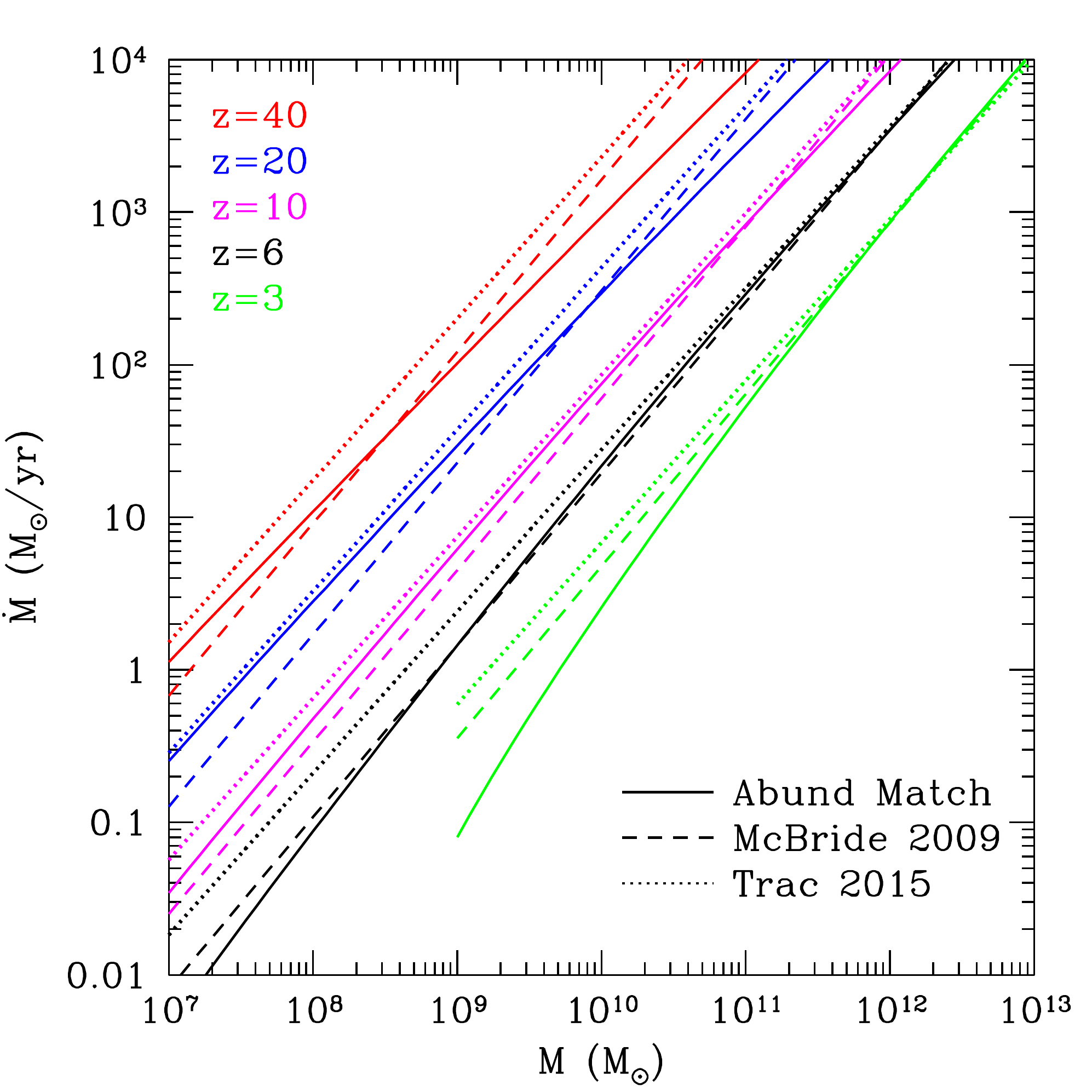}
    \caption{Total accretion rate as a function of halo mass, for $z=3,\,6,\,10,\,20,$ and 40 (from bottom to top). The solid lines take our fiducial abundance-matched accretion rates (from eq.~\ref{eq:halo-mod-acc}). We compare these to the fits from \citet{mcbride09} (dashed lines) and from \citet{trac15} (dotted lines).  The former is based on numerical simulations of moderate to large halos at $z \la 6$ (note the close agreement between the two approaches at $z=3$ for $M \ga 10^{11} \ M_\odot$), while the latter is based on numerical simulations at $z \sim 6$--$10$. }
    \label{fig:acc-rate}
\end{figure}

Figure~\ref{fig:acc-rate} compares these two approaches across a broad range of redshifts and halo masses.  The solid lines show results from abundance matching, while the other lines show simulation-based fits following equation~(\ref{eq:sim-gen}).  The dashed lines take $\mu=1.127$ and $\beta=5/2$ \citep{mcbride09}, while the dotted lines take $\mu=1.06$ and $\beta=5/2$.  We show results for several different redshifts: from bottom to top, the sets of curves correspond to $z=3,\,6,\,10,\,20$, and 40. Although none of these models is likely to be correct in detail, given the inevitable stochastic growth rates of halos, it is reassuring that they match so well without any tuning required. The approaches disagree at small masses at $z \sim 3$, where the simulation fit has not been tested and where abundance matching is complicated by the behavior of the mass function,\footnote{In particular, our model fails at the low-mass end when halos of a given mass have a declining number density, due to their incorporation into more and more massive halos.} but they match very well at $M \ga 10^{10.5} \, M_\odot$.  The agreement is excellent in the range $6 \la z \la 10$, but it worsens at higher redshifts. This is not surprising: the mass functions themselves have not been carefully tested at $z \ga 10$, and the fit from \citet{trac15} is unlikely to be accurate in this regime.

It is important to note that, although these methods match rather well, the deviations nevertheless cannot be neglected for many applications.  Because the specific mass accretion rate (i.e., $\dot{m}_h/m_h$) is nearly independent of mass, any given halo grows (almost) exponentially with redshift.  \citet{dekel13} show that, neglecting the weak mass dependence in the specific accretion rate and taking the $\mu=5/2$,
\begin{equation}
m_h(z) \approx m_0 e^{-\alpha (z-z_0)} ,
\label{eq:dekel-mass-growth}
\end{equation}
where $\alpha = 0.79$, and $m_0$ is the halo's mass at $z_0$.  The parameter $\alpha$ is proportional to the normalization of the accretion rate, so small deviations in it cause large deviations in the history of individual halos when extrapolated over long time intervals.  

We note here that ongoing accretion onto existing galaxies is not the only route through which stars can form in our model:  as halos cross the star formation threshold $m_{\rm min}$, they convert a fraction of their total baryonic mass to stars.  Because the mass function is so steep at high redshifts, this channel can in principle account for a fair fraction of the total star formation, as we show explicitly in Appendix~\ref{bdryterm}.  Our model ignores the luminosity generated by this phase, because the physics of this process is so uncertain (as is the gas content of the halos, if they were subject to earlier bursts of, e.g., Pop III star formation).  Assuming that the star formation occurs over a fraction of the Hubble time (corresponding to the gas cooling time and/or the dynamical time of the halo), this will not impact our predictions within the observable range, as the star formation efficiency within these systems is so small in our models.

\subsection{Quenching of Accretion} \label{quench}

At large halo masses, semi-analytic models of galaxy formation often appeal to quenching from high virial temperatures and/or AGN feedback to decrease the star formation efficiency in accord with observations. We will include the effects of the virial shock heating as an example here, though it will not have a large effect on most of our results. \citet{faucher11} found that the fraction of gas prevented from cooling onto the central galaxy by the virial shock is\footnote{Obviously, we require that $f_{\rm shock} \le 1$.}
\begin{equation}
f_{\rm shock} \approx 0.47 \left( {1+z \over 4} \right)^{0.38} \left( {m_h \over 10^{12} M_\odot} \right)^{-0.25}.
\label{eq:quench}
\end{equation}
We note that this expression is only approximate, as it resulted from a suite of hydrodynamic simulations without outflows and without metal-line cooling, performed at redshifts smaller than our era of interest. We shall see that it only has a modest effect on the results.

\subsection{Merger-Driven Growth} \label{merger}

A fraction of the gas available to high-$z$ halos arrives in the form of mergers. While this has not been measured directly in the relevant redshift regime, at moderate redshifts (and in more massive halos), the fraction is $\sim 20\%$ \citep{goerdt15}. This induces considerable scatter in the effective accretion rate (and hence instantaneous star formation rate) within halos.  As a contrast to our fiducial assumption of completely smooth growth, we therefore consider an alternative model in which \emph{all} accretion is through major mergers. 

With this assumption, and using the \citet{trac15} accretion rates, we find that the number of major mergers per halo per Hubble time $t_H = H^{-1}(z)$ is 
\begin{equation}
N_{\rm merge} \approx 2.75 \left( {1 + z \over 7} \right),
\label{eq:nmerge}
\end{equation}
where we have neglected the weak mass dependence in the specific accretion rates measured by \citet{trac15}.  If we further assume that each merger is followed by star formation over a timescale equal to the dynamical time of the host halo ($\approx t_H/\sqrt{\Delta_{\rm vir}}$, where $\Delta_{\rm vir} = 18 \pi^2$ is the virial overdensity), the fraction of halos actively forming stars at any given time is
\begin{equation}
f_{\rm merge} \approx 0.2 \left( {1+z \over 7} \right).
\end{equation}
Interestingly, these high-$z$ halos are growing extremely fast, so that even in this extreme picture each one is forming stars for a large fraction of the Universe's history.  (Clearly this picture breaks down at sufficiently early times, when $f_{\rm duty} > 1$, but we will not concern ourselves with such early star formation here.)

To model this possibility, we assume that a fraction $f_{\rm merge}$ of halos at each mass are actively forming stars, but during those episodes they accrete gas at a rate $1/f_{\rm merge}$ larger than the value provided by our abundance-matching prescription.

\section{Feedback-Regulated Star Formation} \label{sf-model}

We make the simple ansatz that the star formation rate in a galaxy, $\dot{m}_\star$, is a balance between gas accretion and stellar feedback, which could arise from radiation pressure, supernovae, or some other process like grain heating. We therefore write
\begin{equation}
\dot{m}_{b} = \dot{m}_\star + \dot{m}_w,
\label{eq:halo-growth}
\end{equation}
where $\dot{m}_{b}$ is the rate at which baryons accrete onto the halo (which we assume to be $\dot{m}_b = [\Omega_b/\Omega_m] \dot{m}_h$) and $\dot{m}_w$ is the rate at which baryons are expelled. We then define the star formation efficiency as $f_\star =  \dot{m}_\star / \dot{m}_{b}$ and assume that the rate at which gas is expelled by feedback is proportional to the star formation rate, so that $\dot{m}_w = \eta \dot{m}_\star$, where in general $\eta$ can be a function of halo mass and redshift. Then
\begin{equation}
f_\star = { f_{\rm shock} \over 1 + \eta(m_h,z)} .
\label{eq:fstar-eta}
\end{equation}
Here we have inserted $f_{\rm shock}$ into the numerator in order to include the suppression of accretion by the virial shock; our $f_\star$ therefore represents the fraction of gas that is transformed into stars, relative to that which one naively assumes to accrete onto a halo.

This model is certainly a simplification. For example, it ignores gas residing in the galaxy's interstellar medium (ISM), but if that component grows in proportion to the halo as well, then it will not affect the later arguments substantially.\footnote{In particular, if we assume that the ISM grows at a rate proportional to $m_\star$, we simply change the constant term in the denominator of eq.~(\ref{eq:fstar-eta}). If instead the ISM growth rate is proportional to $\dot{m}_{b}$, we simply change the overall normalization of $f_\star$. Both of these aspects are already quite uncertain.}  

The key assumption so far is that there is no limitation on the star formation rate other than feedback. Because our expression for $f_\star$ is an increasing function of halo mass, this is certainly a poor assumption at high enough halo masses, where $f_\star$ eventually approaches unity.  In the calculations that follow we therefore impose a maximum $f_{\star, {\rm max}}$, which for now we leave as a free parameter. Our results below will depend on the derivative of $f_\star$ with respect to mass, so we impose this maximum through
\begin{equation}
f_\star = {f_{\rm shock} \over f_{\star,{\rm max}}^{-1} + \eta(m_h,z)}
\end{equation}
in order to maintain continuity.

Note that our model specifies the \emph{instantaneous star formation efficiency}, or the fraction of accreting gas converted into stars.  The total star formation rate will be $f_\star \dot{m}_b$: even though our prescription applies equally well to smooth accretion and merger events, the latter will have higher overall star formation rates at fixed $f_\star$ because of their increased (temporary) accretion rates.  

\begin{figure}
	\includegraphics[width=\columnwidth]{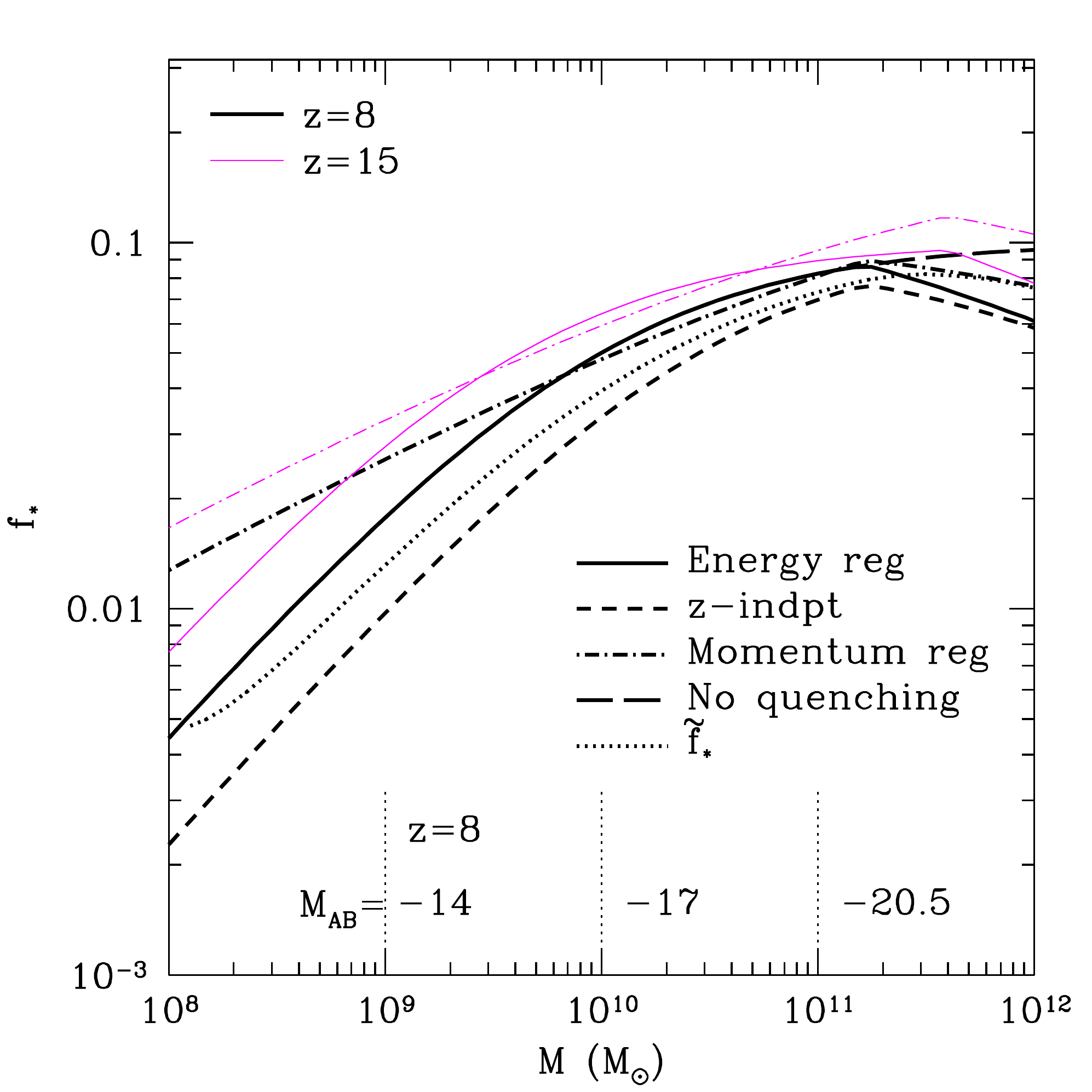}
    \caption{Star formation efficiency $f_\star$ in several models providing reasonable fits to the observed luminosity functions. The thick (thin) curves show $f_\star$ at $z=8$ (15). The solid curves use our energy-driven wind model with $\epsilon_K=0.1$ and $f_{\star,{\rm max}}=0.1$.  The dot-dashed curves use our momentum-driven wind model, with $\epsilon_p=0.2$ and $f_{\star,{\rm max}}=0.2$. The long-dashed line ignores virial shock quenching in the energy-driven model. The short-dashed line takes equation~(\ref{eq:fb-general}) with $\xi=2/3$ and $\sigma=0$ (i.e., assuming it is independent of redshift). The dotted curve shows $\tilde{f}_\star$, or the average star formation efficiency over a halo's history, in our energy-driven wind model at $z=8$.  Along the bottom axis we also show how these masses translate into absolute magnitudes in our energy-driven wind model at $z=8$.}
    \label{fig:fstar}
\end{figure}

\subsection{Models of Feedback} \label{fb-models}

In our fiducial model, we obtain $\eta$ by assuming that the star formation rate is set by the balance between feedback and the forces binding the gas to a galaxy.  However, the coupling between the feedback source and the galaxy's ISM is not yet clear, so we parameterize the feedback mechanism in a flexible manner.

One scenario appeals to supernovae to control the star formation rate.  The most extreme case assumes that supernova blastwaves retain their energy long enough to disturb the accreting gas, so that we can balance the rate of energy input from supernovae with the rate at which the accreting gas acquires binding energy.  Then
\begin{equation}
{1 \over 2} \dot{m}_w v_{\rm esc}^2 = \dot{m}_\star \epsilon_K \omega_{\rm SN},
\label{eq:eta-defn}
\end{equation}
where $v_{\rm esc}$ is the halo escape velocity, $\omega_{\rm SN} = 10^{49} \omega_{49}$~erg $M_\odot^{-1}$ is the energy released in supernovae per unit mass of star formation (determined by the stellar IMF and metallicity), and $\epsilon_K$ is the fraction of that energy released in the wind. Here $\omega_{49}$ is of order unity for a typical IMF.  This sets our fiducial feedback parameter to be
\begin{equation}
\eta_E = 10 \epsilon_K \omega_{49} \left( {10^{11.5} \, M_\odot \over m_h } \right)^{2/3} \left( {9 \over 1+z} \right).
\label{eq:eta-fiducial}
\end{equation}

The prescription in equation~(\ref{eq:eta-defn}) assumes that a fixed fraction $\epsilon_K$ of the supernova kinetic energy is available to lift gas out of the dark matter halo. In fact, the high densities and temperatures of supernova blastwaves at these early times, as well as collisions between nearby ejecta, imply that some fraction of the energy will be lost to radiative cooling or other processes. If this fraction is large, feedback is much less efficient. A more conservative limit on the star formation rate is therefore provided by momentum conservation. We compare the momentum released in supernovae (or other feedback mechanisms, like radiation pressure) to the momentum required to lift the gas out of the halo at the escape velocity.  We write the momentum injection rate as
\begin{equation}
\dot{P} = \pi_{\rm fid} \dot{P}_0 \left( {\dot{m}_\star \over M_\odot /{\rm yr}} \right),
\label{eq:pdefn}
\end{equation}
where $\pi_{\rm fid}$ is of order unity for a typical IMF and $\dot{P}_0 = 2 \times 10^{33}   \ \rm{g \, cm/s}^2$ (which equals the momentum input from a Salpeter IMF with solar metallicity).  If a fraction $\epsilon_p$ of this momentum is used to drive a wind, we have $\epsilon_p \pi_{\rm fid} \dot{P}_0 \dot{m}_\star = \dot{m}_b v_{\rm esc}$, or 
\begin{equation}
\eta_p = \epsilon_p \pi_{\rm fid} \left( {10^{11.5} \, M_\odot \over m_h } \right)^{1/3} \left( {9 \over 1+z} \right)^{1/2}.
\label{eq:etap_fiducial}
\end{equation}

Because feedback transitions between these two regimes, and because other physics is most certainly relevant as well, we will allow for a more general form and parameterize the feedback efficiency as a power law in mass and redshift,
\begin{equation}
\eta = C \left( {10^{11.5} \, M_\odot \over m_h } \right)^{\xi} \left( {9 \over 1+z} \right)^{\sigma},
\label{eq:fb-general}
\end{equation}
where $C$ is a normalization constant that can be fixed by comparison to observations for a given set of power law indices. \citet{sun16} found that $\xi \sim 1/3$--2/3 (bracketed by our energy and momentum conservation models) provides an adequate fit to the observed luminosity functions at $z \ga 6$, but the redshift evolution is less well-quantified and other mass dependence is certainly allowed.

Figure~\ref{fig:fstar} shows the star formation efficiency in several models that provide adequate fits to the existing observational data, (see below). The thick and thin curves show the results at $z=8$ and 15, respectively.  The solid lines take our fiducial energy-regulated wind model with $\epsilon_K=0.1$ and $f_{\star,{\rm max}}=0.1$, typical parameters for a supernova wind model.  Note that, at small masses, $f_\star$ increases with redshift because the binding energy of the host halo does as well, allowing more star formation before the wind breaks out.  The dot-dashed curves show the momentum-regulated wind model with $\epsilon_K=0.2$ and $f_{\star,{\rm max}}=0.2$. The short-dashed line shows a version motivated by the mass scaling of the energy-regulated feedback, with $\xi=2/3$, but with no redshift dependence ($\sigma=0$).  The overall normalization corresponds to $\epsilon_K = 0.2$ at $z=8$. A larger value of $\epsilon_K$ decreases the overall star formation rate because more of the supernova energy is used to drive winds. Finally, the long-dashed line takes our fiducial energy-driven wind parameters but ignores quenching from virial heating, which only affects the largest masses.  (It is this quenching that imposes the kink in $f_\star$ above $10^{11} \ M_\odot$.)

Along the bottom axis of Figure~\ref{fig:fstar}, we also show the absolute magnitudes corresponding to the solid curve at $z=8$ (see Section \ref{lum-conv} for details).  Surveys with substantial volumes currently reach $M_{\rm AB} \approx -18$ at this redshift, so current observations only probe halo masses near the peaks of these curves.  This illustrates why a physical model to guide extrapolation to fainter luminosities is so important.

\subsection{The integrated star formation efficiency} \label{int-sf}

Our star formation efficiency parameter $f_\star$ describes the fraction of accreted gas turned to stars over a short timescale (formally, instantaneously). This definition differs from a more common form in the literature, which we will label $\tilde{f}_\star$. This is the fraction of the baryons associated with a halo that has been transformed into stars:\footnote{Note that this is also not quite the fraction of a halo's nominal baryonic mass that is in stars at any given time, because some fraction of the stars would have already completed their core fusion life cycles. }
\begin{equation}
{m_\star \over m_h} = \tilde{f}_\star {\Omega_b \over \Omega_m}.
\end{equation}

We can compare $f_\star$ and $\tilde{f}_\star$ through direct integration of the accretion rate and star formation rate for each halo. Unfortunately, even in our star formation prescription, there is no simple expression for that relation.  However, we can obtain an approximate form by using equation~(\ref{eq:sim-gen}) with $\mu=1$ (i.e., a mass-independent specific accretion rate), assuming that the accreted mass is much larger than the initial mass, taking equation~(\ref{eq:fb-general}) with $\eta \gg 1$ for the feedback efficiency, and ignoring $f_{\star, {\rm max}}$. Then
\begin{equation}
\tilde{f}_\star(m_h,z) \approx {f_\star(m_h,z) \over 1 + \xi}.
\label{eq:ftildestar-approx}
\end{equation}
For our energy-driven and momentum-driven models, this yields $\tilde{f}_\star \approx (0.6$--$0.75) f_\star$. 

Figure~\ref{fig:fstar} shows this integrated efficiency at $z=8$ in our fiducial energy-driven wind model with the dotted curve.  Our approximation is reasonably accurate except for the smallest masses (where we have assumed a pre-existing mass of stars equal to the instantaneous $f_\star$) and near the peak efficiency (where quenching and $f_{\star,{\rm max}}$ introduce complexities). 

We emphasize that both versions have (approximately) the same mass and redshift dependence; because the normalization factor is uncertain by at least a factor of two, the distinction between $f_\star$ and $\tilde{f}_\star$ is likely not terribly important in comparison to observations. The net star formation efficiency is smaller than the instantaneous version because $f_\star$ is an increasing function of mass (which, for any given halo, is growing exponentially with time, outweighing the power-law increase in the star formation efficiency at a fixed mass at higher redshifts).

\subsection{From Star Formation to Luminosity } \label{lum-conv}

Finally, in order to compare our model to observations we must translate our star formation rates into luminosities.  Because the UV luminosity of (non-active) galaxies results from massive, short-lived stars, it is a good tracer of the star formation rate.  We there take the standard conversion
\begin{equation}
\dot{m}_\star = {\mathcal K}_{\rm UV} \times L_{\rm UV}
\label{eq:lum-conv}
\end{equation}
where $L_{\rm UV}$ is the intrinsic (i.e., without dust) luminosity in the rest-frame continuum (1500--2800 \AA) and ${\mathcal K}_{\rm UV}$ is a proportionality constant that depends on the IMF, star formation history, metallicity, binarity, etc. Following \citet{sun16}, we take ${\mathcal K}_{\rm UV} = 1.15 \times 10^{-28} \ M_\odot$ yr$^{-1}$ / (ergs s$^{-1}$ Hz$^{-1}$), which assumes a low-metallicity Salpeter IMF with continuous star formation.  We note that uncertainty in this conversion affects the overall amplitude of the star formation efficiency, which is therefore uncertain by a factor of a few.

Note that we ignore dust, both in our abundance matching procedure and in our luminosity estimates from the star formation rate.  Based on recent measurements, the galaxy dust correction appears to be declining rapidly in the $z \sim 6$ regime \citep{dunlop13,bouwens14}.  Bright galaxies have some evidence for dust at this time, but fainter ones do not appear to require it.  At even higher redshifts, there is no evidence for dust even in the bright populations.  As a result, other studies that have included dust find it provides only a modest increase to the estimated star formation rates at $z > 6$ (see, e.g., Fig.~7 of \citealt{sun16} and Fig.~10 of \citealt{mason15}) except in the brightest galaxies. Nevertheless, we have neglected it here because we have found that, in this limit where the dust  is evolving rapidly, the correction depends sensitively on its parameterization and requires careful consideration (see \citealt{smit12} for the standard method). In the spirit of our very simple treatment, we have ignored it here, but we note that it could affect our interpretation of the bright end of the galaxy luminosity function. These uncertainties about parameterizing the dust content have also provided one of our motivations for confining our comparisons to $z \geq 6$.

\section{Results} \label{sfrhist}

The simple tools in the preceding section allow us to compute model star formation histories of the Universe.  In this section, we compare our results to the galaxy populations at $z \ga 6$.  

\begin{figure}
	\includegraphics[width=\columnwidth]{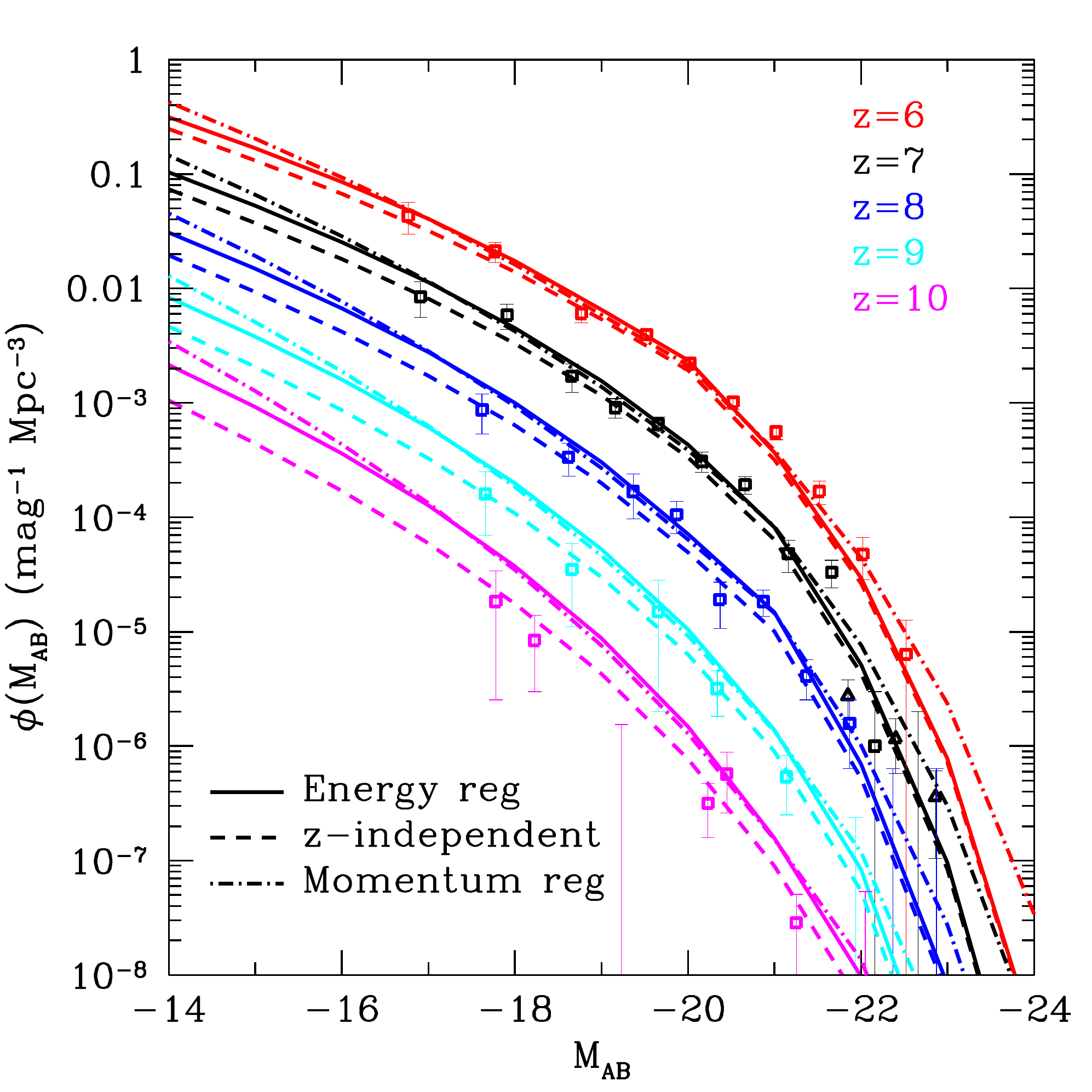} 
    \caption{Comparisons of our model luminosity functions with the data at $z=6$--10, including our fiducial energy-regulated model, momentum-regulated model, and redshift-independent model (solid, dot-dashed, and dashed curves, respectively).  See the text for detailed parameter choices. In each panel, the $z=6,\,7,\,8,\,9$, and 10 luminosity functions are displaced by $+0.5,\,0,-0.5,-1,$ and -1.5 in logarithmic units along the ordinate for ease of display. We compare to luminosity functions from \citet{bouwens15}, supplementing with data from \citet{oesch13} and \citet{bouwens15b} at $z=9$--10 and from \citet{bowler16} at $z =7$. }
    \label{fig:lfcomp}
\end{figure}

\subsection{Comparison to measured luminosity functions} \label{lfcomp}

We first consider whether our simple model provides a reasonable fit to observed luminosity functions at $z \sim 6$--10  and use those observations to calibrate the free parameters in each model (the efficiency of feedback, $\epsilon_K$ or $\epsilon_p$, and the maximal star formation efficiency $f_{\star,{\rm max}}$)\footnote{In detail, we compare to luminosity functions measured in bins centered at $z=5.9,\,6.8,\,7.9$ and broad bins at $z \sim 9$ and 10.}. Given the simplicity of our models and the many uncertainties in their implementation, we will not attempt a rigorous statistical test (better suited to more flexible empirical models; see \citealt{sun16,mirocha16}) but will demonstrate under what conditions our models reproduce the existing data, at least roughly.

For concreteness, we will primarily compare our results to luminosity functions measured in the range $z=6$--10 by \citet{bouwens15}, supplemented by \citet{oesch13} and \citet{bouwens15b} at the highest redshifts and \citet{bowler16} at the bright end for $z=7$. We note that several other groups have produced luminosity functions in this range (e.g., \citealt{mclure13, schenker13, finkelstein15}) but we have chosen one group with broad redshift coverage for consistency. The other measurements are generally consistent.\footnote{The \citet{finkelstein15} measurements are somewhat lower in overall number density than the \citet{bouwens15}, but as this amplitude is degenerate with the normalization of the star formation efficiency the difference does not affect our overall conclusions.  The shapes of the luminosity functions are quite similar, as is their redshift evolution.  See \citet{mirocha16} and \citet{mason15} for more explicit comparisons between the two data sets.}

Figure~\ref{fig:lfcomp} compares our model to this data. (Note that we have displaced the luminosity functions in the vertical direction for clarity in presentation, so that they appear with $z=6$ at the top and $z=10$ at the bottom: the overall decline is much smaller than suggested here.) The solid curves show our fiducial energy-regulated model, with $\eta$ set by equation~(\ref{eq:eta-fiducial}).  To roughly match the data, we have set $\epsilon_K=0.1$ and $f_{\star,{\rm max}}=0.1$. These parameters do an excellent job reproducing the data at $z=6$--8, but they lay at the upper end of the allowed range at the higher redshifts. Recall that our energy-regulated model has fairly steep redshift dependence, because the binding energy of halos (at fixed halo mass) increases as $(1+z)$: thus in this picture feedback permits more star formation at higher redshifts.  

To try to better match the different redshifts, the dashed curves use the same mass dependence as the energy-regulated model but assume no redshift dependence (i.e., $\xi=2/3$ and $\sigma=0$ in eq.~\ref{eq:fb-general}). We set the normalization constant to be equivalent to $\epsilon_K=0.2$ at $z=8$ but keep $f_{\star,{\rm max}}=0.1$.  This underpredicts the abundance of moderately bright galaxies at $z=6$ but provides a much better match to the data at $z>8$.

\begin{figure}
	\includegraphics[width=\columnwidth]{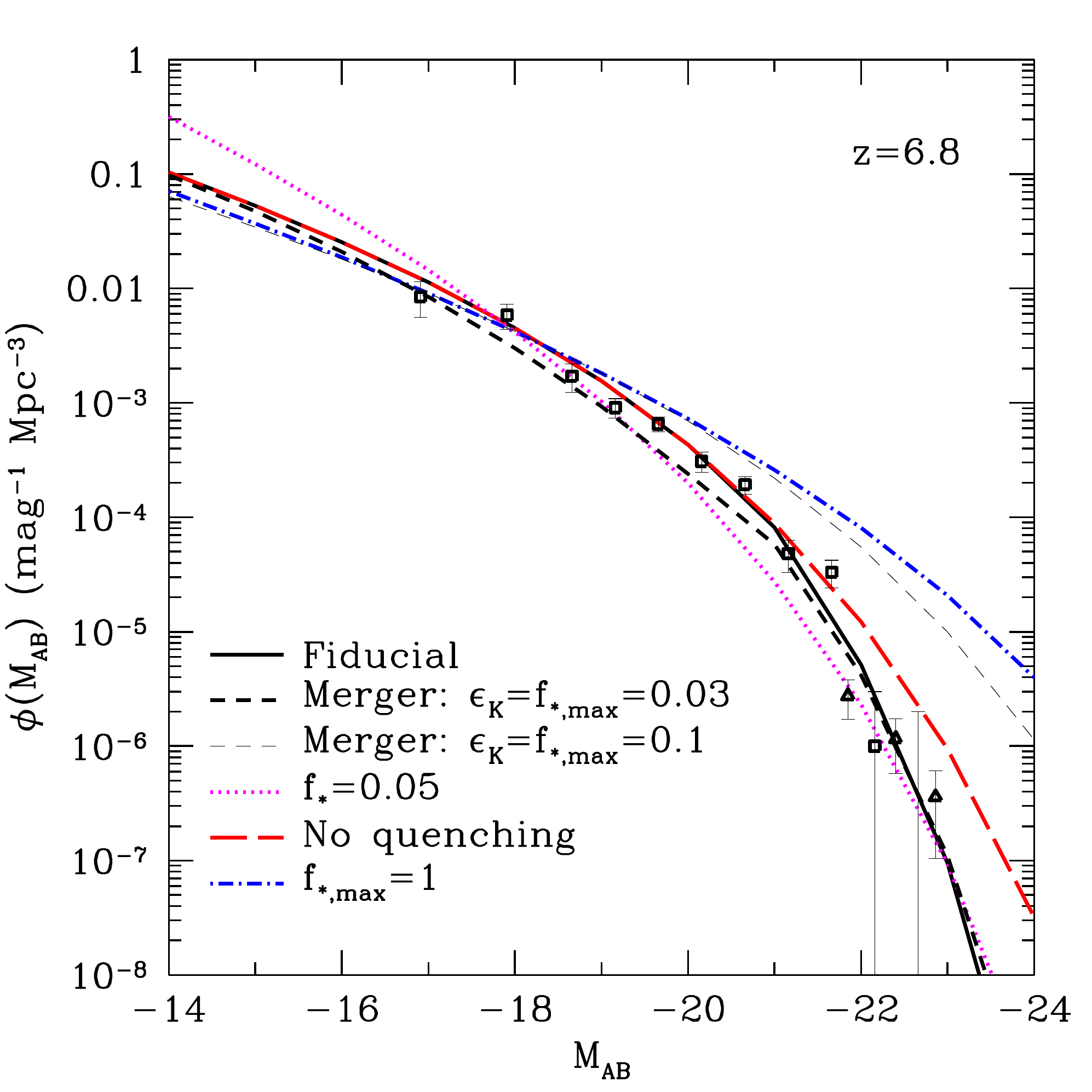} 
    \caption{Illustration of the parameter dependence of our model luminosity functions. We show results at $z=6.8$ and compare to data from \citet{bouwens15} and \citet{bowler16}.  The solid curve is our fiducial energy-regulated model ($\epsilon_K=0.1$ and $f_{\star,{\rm max}}=0.1$). The long-dashed curve ignores quenching of accretion, while the dot-dashed curve also takes $f_{\star,{\rm max}}=1$ and $\epsilon_K=0.2$. The dotted curve assumes a constant $f_\star=0.05$. Finally, the short-dashed curves use our merger model. The thin, short-dashed curve follows our fiducial parameters, without quenching, while the thick short-dashed curve provides a better fit to observations by setting $\epsilon_K=0.03$ and $f_{\star,{\rm max}}=0.03$.}
    \label{fig:lfparams}
\end{figure}

The dot-dashed curves use our momentum-regulated model of equation~(\ref{eq:etap_fiducial}), with $\epsilon_p=5$ and $f_{\star,{\rm max}}=0.2$.  This choice provides a reasonably good fit to the data, though it also somewhat overpredicts the amplitude at $z>8$.  Comparison of the momentum and energy regulated models shows that there are enough uncertainties, even in very simple models of galaxies, that at present it is difficult to say anything concrete about the processes regulating star formation at high redshifts. However, differences are stronger at the faint end, where momentum regulation allows for many more faint galaxies, and the redshift dependence is already revealing.

Figure~\ref{fig:lfparams} explores some details of our model.  We focus here on $z=6.8$, comparing to \citet{bouwens15} and \citet{bowler16} with the squares and triangles, respectively. For reference, the solid curve is our fiducial energy-regulated model, with $\epsilon_K=0.1$ and $f_{\star,{\rm max}}=0.1$.  The dotted line assumes a constant $f_\star=0.05$, so that the luminosity function is close to a rescaling of the halo mass function (because $\dot{m}_h$ is nearly proportional to $m_h$ over most of our mass and redshift range; see eq.~\ref{eq:sim-gen} with $\mu \approx 1$). While this roughly reproduces the overall amplitude, it fails badly in reproducing the shape of the luminosity function, demonstrating why models require $f_\star$ to decline towards small halo mass.  

Two other curves show how our parameters affect the bright end.  The long-dashed curve ignores the quenching of accretion: while not an enormous effect at this redshift, some kind of cutoff is nevertheless important to match the bright end.  The dot-dashed curve sets $f_{\star,{\rm max}}=1$ in addition to ignoring quenching: this dramatically overpredicts the number of bright galaxies, indicating that we must impose a saturation limit on the efficiency of star formation within galaxies independent of internal feedback, which becomes ineffective in the very deep potential wells of the most massive halos.  (Note that this model also takes $\epsilon_K=0.2$ to ensure a fit to the faint end.) We also note that these bright galaxies very likely have dust, which we have not modeled.  As it will preferentially extinguish light from the brightest sources, its presence will certainly mitigate (if not eliminate) the need for quenching mechanisms.  For example, \citet{sun16} found, using a standard dust correction, that dust can reduce the UV luminosity of the brightest galaxies at $z \la 8$ by $\sim 2$ magnitudes.  

A particularly interesting result at the bright end is the comparison to the \citet{bowler16} data.  Those authors noted that a double power law provides a better fit to the data than a Schechter function, thanks to the relative abundance of very bright galaxies.  Our models (which are ultimately sourced by the halo mass function) have no trouble fitting the bright end, and in fact -- given the abundance of faint galaxies - demonstrate that, just as at low redshifts, we must impose restrictions on star formation in massive halos in order to reproduce the data.  Thus it is not surprising -- at least from a theoretical perspective -- that a double power law is preferred.  

The other two curves in Figure~\ref{fig:lfparams} explore our merger prescription.  First, the thin short-dashed curve uses our fiducial parameters from the energy-regulated model ($\epsilon_K=0.1$ and $f_{\star,{\rm max}}=0.1$), but assumes that all star formation occurs in merger events.  This dramatically overpredicts the abundance of bright galaxies.  Recall that our extreme merger picture has the same \emph{average} accretion rate as the smooth models, but it assumes that the gas inflow is large during brief bursts and off otherwise.  Thus, when a galaxy is active, it has a much higher total star formation rate.  Using our fiducial parameters therefore overpopulates the bright end by shifting smaller halos toward larger luminosities.  A better fit can be obtained by reducing the maximal efficiency of star formation while also decreasing the importance of feedback ($\epsilon_K=0.03$ and $f_{\star,{\rm max}}=0.03$).  

Recently, \citet{oesch16} discovered a very bright galaxy ($M_{AB} \sim -22.1$) with an inferred redshift of $z \sim 11.1$.  Unless this object proves to be a fluke, it presents a serious challenge to the treatment of very bright galaxies in our models.  Given their survey volume, the apparent number density of sources similar to this is $\sim 8 \times 10^{-7}$~Mpc$^{-3}$, albeit with very large errors from the single object detected.  Our energy-regulated model, the most optimistic of our fiducial choices, falls about a factor of ten short of this inferred number density.  Although this galaxy is very bright, it is not massive enough to be subject to our quenching prescription (eq.~\ref{eq:quench}), and the only way to significantly increase the abundance at this luminosity is to remove the saturation level of star formation.  We also note a tension with data from slightly lower redshifts:  recall that our fiducial models already tend to overpredict the galaxy number density at $z \sim 9$--10.  Adjusting our parameters to fit this $z \sim 11.1$ galaxy would dramatically increase the discrepancy at slightly lower redshifts: indeed, our redshift-independent model, which fits the $z \sim 9$--10 data nicely, is strongly discrepant from the $z=11$ object.  For now, this object remains very puzzling, though we note that some numerical simulations have found very bright objects to be more common than our models predict (see below; \citealt{waters16,mutch16b}).

Importantly, we find that a variety of feedback and merging prescriptions can reproduce the luminosity functions over the redshift interval $z \sim 6$--10.  Thus the data cannot yet discriminate between these physical models, though there are hints of differences.  Nevertheless, we shall see in the next section that our general framework still makes strong qualitative predictions for the evolution at higher redshifts.

\begin{figure*}
	\includegraphics[width=\columnwidth]{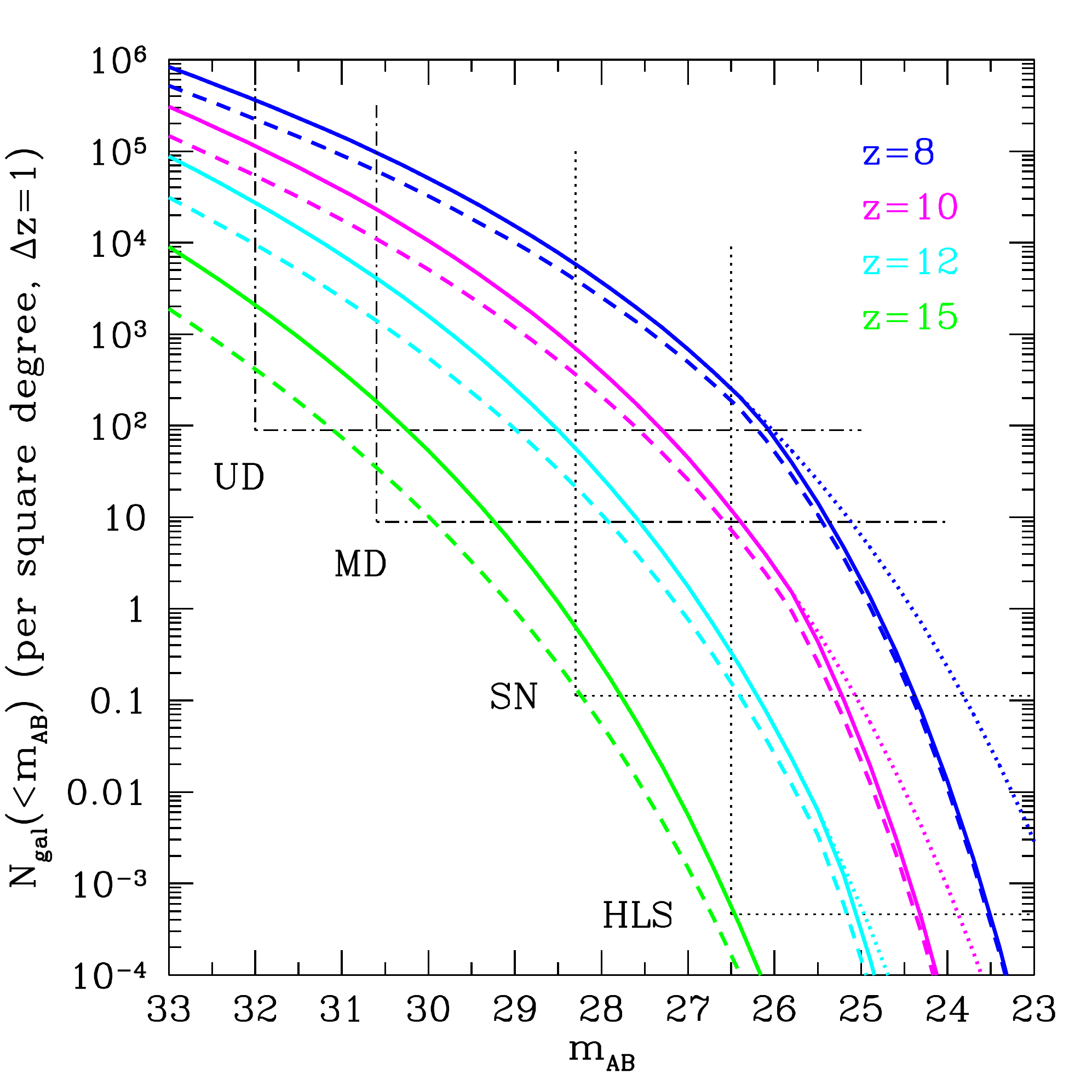} \includegraphics[width=\columnwidth]{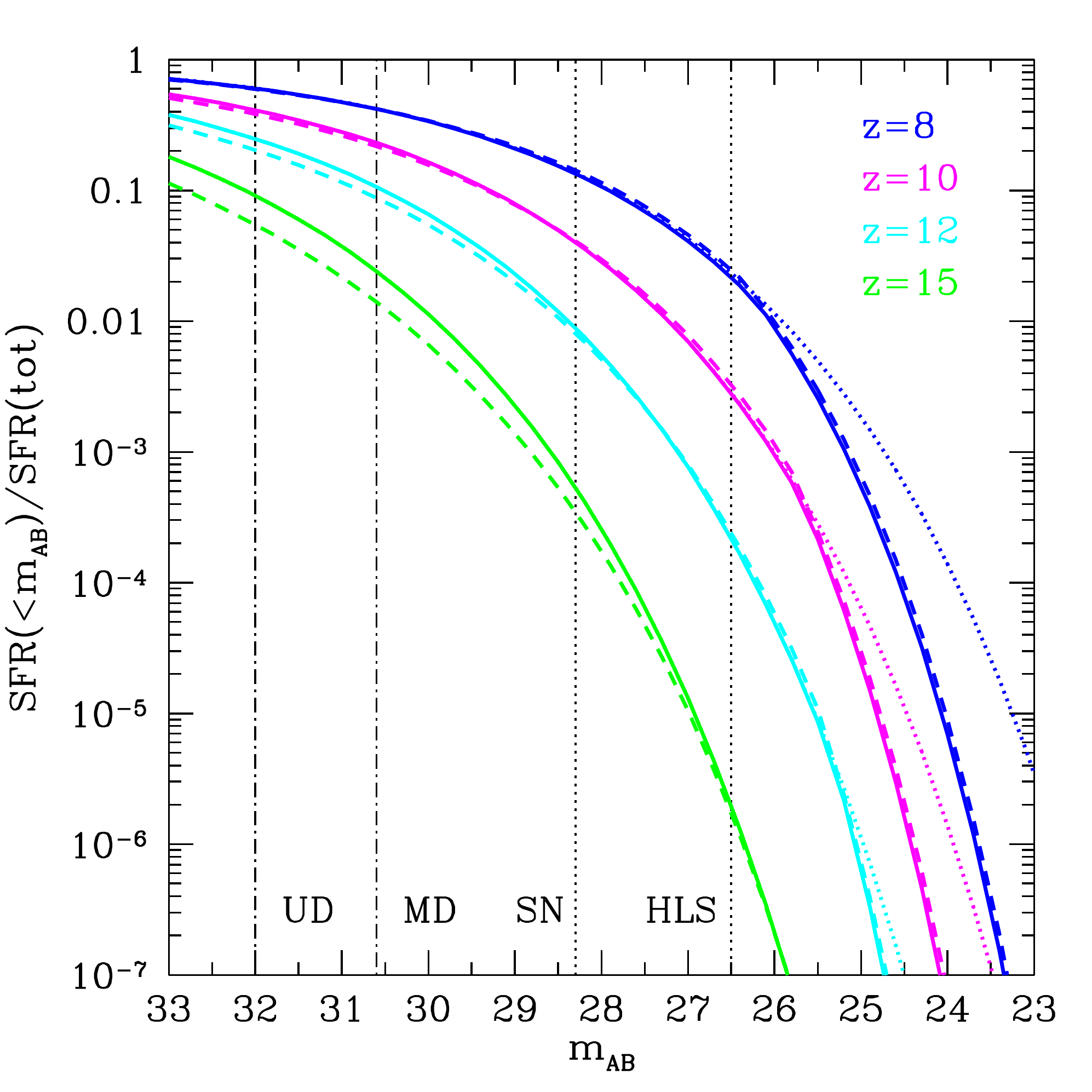}
    \caption{\emph{Left:} Surface density of high-$z$ galaxies in our models, with comparisons to several potential surveys.  The curves show the number of galaxies more luminous than the specified apparent magnitude (in AB units) per square degree in surveys spanning $(z-0.5,z+0.5)$, where $z=8,\,10,\,12,$ and 15, from top to bottom.  Within each set, the solid and dashed curves show our fiducial energy-regulated and redshift-independent models.  The dotted curves use our energy-regulated model but ignore quenching.  We compare these surface densities to four potential space-based surveys. The UD and MD surveys each assume 800 hour total survey time with JWST, while SN and HLS refer to surveys similar to those in WFIRST's cosmology program. In each case, we show the approximate limiting magnitude (vertical lines) and the inverse of the survey's area. \emph{Right:} Fraction of total star formation inside galaxies brighter than the specified limiting magnitude.  The curves are identical to those in the left panel. We assume that star formation extends to the atomic cooling threshold at all redshifts.  Limiting depths of surveys are indicated by vertical lines.}
    \label{fig:lf_highz}
\end{figure*}

\subsection{Future surveys} \label{lfexp}

One of the key advantages of a physical model for galaxy formation is its utility in forecasting the results of future surveys. While such predictions invariably prove wrong, they provide a framework for planning survey depths, areas, and strategies that can make substantial progress over the existing knowledge base.  In particular, our models are calibrated to the \emph{known} galaxy population.  If additional physics or objects are present, such as ``minihalos" hosting Population~III stars, they will be absent from our forecasts.  Our model essentially provides a conservative extrapolation of the existing data, and the most interesting result from future surveys will be to discover deviations from these expectations.

To develop a baseline prediction, we next use our models to estimate the number of galaxies that will be observed at a variety of redshifts with two important future instruments, JWST and WFIRST. We consider four fiducial surveys, two with each observatory.  For WFIRST, we include the High Latitude Survey (HLS), which will span 2227 square degrees to an approximate infrared depth of $m \approx 26.5$, as well as a nine square degree deep field reaching a depth of $m \approx 28.3$.  Both of these surveys approximate core parts of the mission that are structured for cosmological observations (lensing, baryon acoustic oscillations, and supernovae) but will prove useful for many other science questions.  For JWST, we follow \citet{mason15} and consider an ``ultradeep" survey consisting of four 200-hour pointings (each $\approx 10'$), which reaches $m \approx 32.0$, and another ``medium deep" survey consisting of forty 20-hour pointings, which reaches $m \approx 30.6$.\footnote{\citet{mason15} calculated these depths using the JWST Exposure Time Calculator in its pre-Cycle 1 form. They also considered a wide survey with similar depth to the WFIRST SN survey, but with only about 1/5 the area.}  Note that we do not attempt to model the detailed selection functions or individual filter depths of any of these potential surveys, as the JWST programs are purely hypothetical, while WFIRST has not yet settled on a detailed mission design.  We also neglect gravitational lensing, which can substantially affect the bright end \citep{mason15}.  Also, note that WFIRST will only have coverage to $\approx2$~$\mu$m, so its maximum redshift will be limited.

The left panel of Figure~\ref{fig:lf_highz} shows our model predictions for the surface density of galaxies from $z=8$--15, assuming surveys spanning $(z-0.5,z+0.5)$.\footnote{We assume a flat spectrum in $f_\nu$ to estimate the $K$-correction for the AB magnitude flux density.}  We show our fiducial energy-regulated model with the solid curves and the redshift-independent model with the dashed curves; the former is representative of a more optimistic extrapolation to higher redshifts, while the latter is our most pessimistic prescription.  Because our model may underpredict the number of very bright objects (see section \ref{comp-model} below), we also show our energy-regulated model with quenching turned off (dotted curves).

We compare these predictions to the four surveys outlined above. We illustrate the parameter space each can probe by marking their AB limiting magnitudes and the inverse of their areal coverage (which corresponds to the surface density above which they are expected to detect \emph{at least one} object).  From top left to bottom right, these are the ultradeep (UD), medium deep (MD), supernova (SN), and HLS programs.  

In our models, the characteristic luminosity declines rapidly with redshift. By $z \sim 15$, objects bright enough to be seen by the shallowest surveys are very, very rare.  Thus it appears that, even with future instruments, extremely deep surveys will be required to detect objects at $z \ga 15$.  Even then, our UD survey would detect, at best, a few dozen objects at $z \sim 15$. The solid and dashed curves in this figures roughly correspond to optimistic and pessimistic extrapolations from models calibrated to fit the observed data.  The largest differences, at the faintest luminosities and highest redshifts, are up to an order of magnitude, which shows the precision required to make useful constraints on the physics of galaxy formation during this era (and/or detect deviations from our models expectations).  

However, all four of these surveys would detect many thousands of galaxies at $z \sim 8$-12.  None will span the entire range of the luminosity function, but in combination they can map out a dynamic range of up to eight magnitudes.  As an additional, albeit crude, figure of merit, in the right panel of Figure~\ref{fig:lf_highz}, we show the fraction of the total star formation at each redshift occurring in halos brighter than the specified limiting magnitude.  In all cases, we assume that star formation extends to the minimum mass for atomic cooling.  Because of the steepness of the luminosity function,the deeper surveys are better at probing the majority of the emission: our UD survey, for example, is sensitive to $\sim 60\%$ of the total star formation at $z=8$.  That fraction falls to $\sim 10\%$ at $z=15$, but that is still far more than the other surveys.  

Nevertheless, we note that even JWST will require either large lensing surveys or substantial extrapolation to fainter luminosities if we are to compute quantities depending on the \emph{total} star formation rate density in the universe (such as reionization).  We expect that precision constraints on the shape of the luminosity function will improve substantially through the improved statistics, which will be essential for extrapolating to the faint galaxies responsible for most of the emission.  In this respect, large-area surveys will be extremely useful in avoiding biases due to cosmic variance \citep{robertson10}.  They are also essential for probing the full range of galaxy luminosities in order to study how galaxies evolve in this early epoch (for example, measuring how chemical enrichment occurs as galaxies grow) and for cross-correlation with, e.g., highly-redshifted 21-cm measurements and intensity mapping, both of which require very large spatial volumes \citep{lidz09,lidz11}.

\begin{figure}
	\includegraphics[width=\columnwidth]{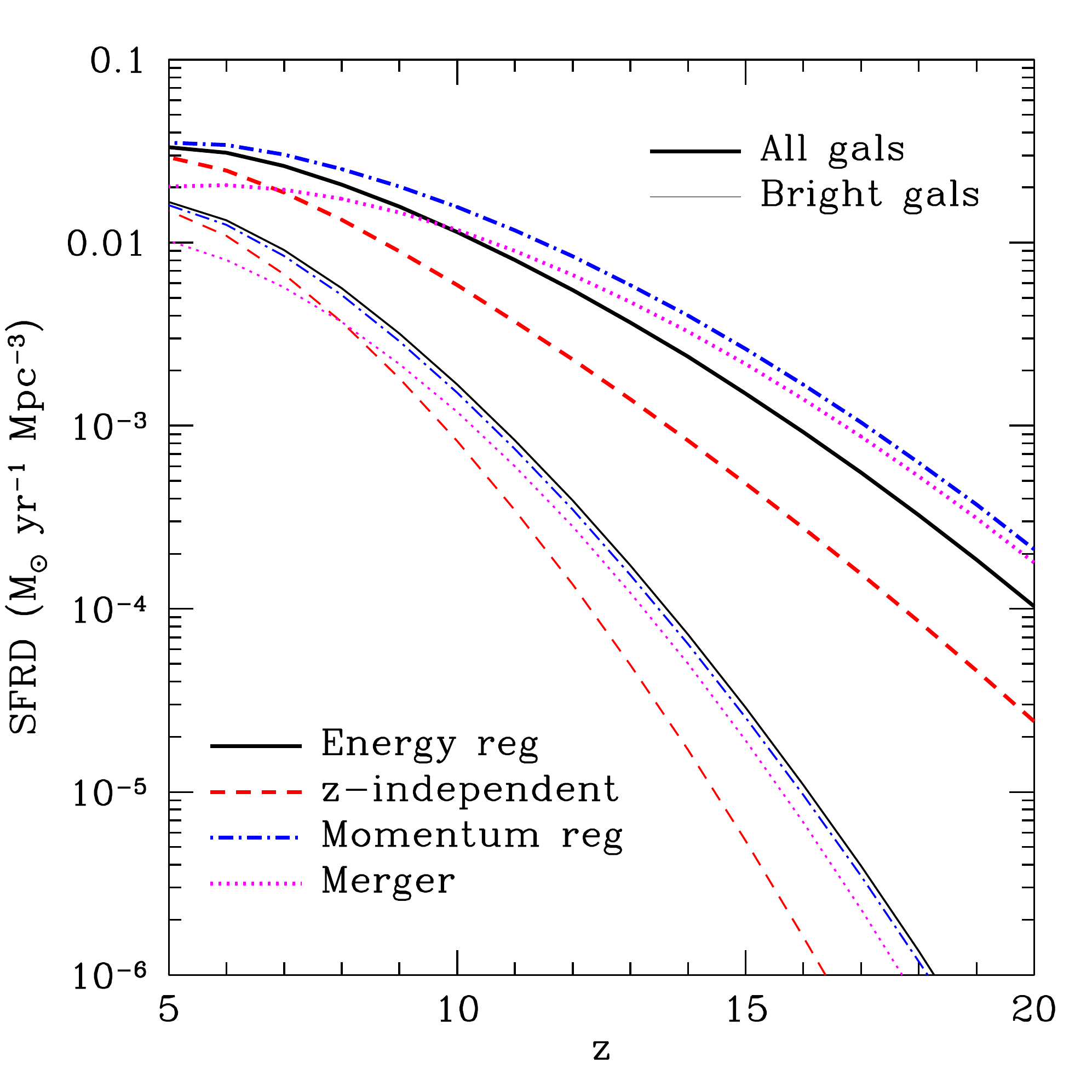} 
    \caption{Integrated star formation histories of our models. The upper set of curves include galaxies down to the minimum virial temperature for atomic cooling, $10^4$~K.  The lower set of thin curves include only galaxies with $M_{\rm AB} < -17.7$.  Within each set, the solid, dot-dashed, dotted, and dot-dashed curves use our energy-regulated, merger-driven, momentum-regulated, and redshift-independent models, respectively. }
    \label{fig:sfhistory}
\end{figure}

\subsection{The star formation history} \label{sfhist}

We next consider the evolution of the integrated star formation rate (and hence UV emissivity) over a broad redshift range, as shown in Figure~\ref{fig:sfhistory}.  Here the upper set of curves show the total star formation rate density, assuming that all halos above the atomic cooling threshold (at a virial temperature $T_{\rm vir} \approx 10^4$~K) are able to form stars.  Three of our models show very similar behavior: these are the energy-regulated feedback model (solid curve), the  momentum-regulated model (dot-dashed curve), and  energy regulation with mergers (dotted curve).  Although these have different physical assumptions, calibrating them to the observed luminosity functions at $z \sim 6$--8 leads to very similar integrated star formation histories, within a factor $\sim 2$ of each other over the entire redshift range.  In detail, the momentum-regulated case has more star formation at most redshifts (because it imposes a weaker constraint on small halos), while the merger-driven model falls short of the others at $z \la 10$ (because it underpredicts the abundance of moderate-luminosity galaxies; see Fig.~\ref{fig:lfparams}). The dashed curve is the same redshift-independent model shown in Figure~\ref{fig:lfcomp}, which fits the $z\sim 10$ data better.  It decreases more steeply toward early times, because (unlike the other models) the star formation efficiency of small halos does not increase at higher redshifts. 

The lower set of thin curves of Figure~\ref{fig:sfhistory} shows the star formation rate density in bright galaxies, here including all those with $M_{AB} < -17.7$, as often calculated from observations.  Unsurprisingly, this evolves much more rapidly than the total star formation rate density, as halos massive enough to host bright galaxies become exponentially more rare toward early times.  This is also clear from the right panel of Fig.~\ref{fig:lf_highz}: at $z=8$ (15), $M_{AB}=-17.7$ corresponds to $m_{AB}\approx 29.5$ (30.4).  All of our models have similar behavior in this respect.

For completeness, we also note that, although we have allowed star formation to persist to very small halo masses ($\sim 10^8 \ M_\odot$), at $z \sim 10$ the smallest halos produce a smaller fraction of the star formation than in many previous treatments: at $z=8$, halos with $M \ga 10^9 \ (10^{10}) \ M_\odot$ produce $\ga 75$\% (30\%) of the total star formation (see the right panel of Fig.~\ref{fig:lf_highz}).  This is because, in our model, internal feedback strongly suppresses star formation in shallow potential wells.

\subsection{Comparison to existing models} \label{comp-model}

We next consider our approach in the context of the many other models of the high-$z$ galaxy luminosity function.  Most obviously, our approach relies on a theoretically-motivated framework to model the galaxy populations rather than providing empirical fits to the measurements. Several groups have used empirical fits to the high-$z$ luminosity functions, with a range of parameterizations, to extrapolate the results to higher redshifts. \citet{robertson13, robertson15} and \citet{oesch13} provided fits to the global star formation rate. Their results demonstrate the difficulty of empirical extrapolations, as the extrapolations are dramatically different because of different assumptions about the continuity of the imposed functional form.  Of course, more detailed observations provide more basis for such empirical extrapolations, and the improving measurements of the galaxy luminosity functions in this high-$z$ regime have led to several recent attempts at more sophisticated empirical fits (e.g., \citealt{mason15, visbal15, sun16, mirocha16}).  All of these models agree that $f_\star(m)$ peaks at $m \sim 10^{11} \, M_\odot$ and declines toward smaller masses, but they are also subject to systematic uncertainties in the way in which that function is parameterized (for example, whether or not redshift-dependence is allowed).  \citet{sun16} and \citet{mirocha16} included error estimates on their fit parameters, and our example models are all within the bounds allowed by those estimates.

The key difference between our approach and these other works is our reliance on an underlying theoretical model.  There are clear advantages to both methods.  Our theoretical framework has three basic ingredients: the dark matter halo mass function, the halo accretion rate, and the feedback model. To the extent that these fundamental ingredients are trustworthy, our model provides more reliable extrapolations to higher redshifts, because it ``builds in" the proper redshift evolution.  We hope that the first two, which depend only on dark matter physics, are reasonably well-understood (and in fact are often built into empirical fits as well).  We have found that the remaining uncertainties in the data allow a fairly wide range of star formation prescriptions (in our case, expressed through the feedback model), but that normalizing to the $z \la 10$ data nevertheless significantly limits the range of extrapolated star formation densities.   Moreover, we have found that the bright galaxy population becomes an increasingly poor proxy for the total star formation rate.  Empirical predictions based on the observed bright galaxies are therefore subject to large systematic uncertainties as they are extrapolated to higher redshifts.

\citet{behroozi15} also used an abundance matching model to predict high-redshift galaxy populations.  Their model, originally developed to describe galaxies at low and moderate redshifts, uses a more complex abundance matching procedure based on stellar mass measurements and accounting for subhalos, scatter in the mass-luminosity relation, and other effects.  Though they model the dark matter physics explicitly, they do not attempt to parameterize the star formation laws as we do.  Because we calibrate our model exclusively at $z \geq 6$, we have focused on the well-measured UV luminosity functions, and we have shown that the data do not yet demand more complex treatments of the halo populations.  Nevertheless, we find qualitatively similar results to \citet{behroozi15}: rapid declines in the halo populations, though an increasing star formation rate with redshift at fixed halo mass.  (In our case, this is primarily accomplished by the rapidly increasing accretion rate, as the star formation efficiency $f_\star$ may or may not increase with redshift.)

\citet{mason15} and \citet{mashian16} use abundance matching to make empirical fits to the star formation efficiency as a function of stellar mass and extrapolate to higher redshifts (see also \citealt{trenti10}).  They include slightly different sets of effects in their modeling than us (for example, \citealt{mashian16} allows scatter in the halo-luminosity relation). Neither finds evidence for redshift evolution in this relation (though they also do not rule it out at the level implied by our model or by \citealt{behroozi15}); however, \citet{mason15} find an increase in \emph{luminosity} with redshift at a fixed halo mass because of the increased growth rate of halos at earlier times, similar to our model's increased accretion rate. \citet{mason15} has the advantage of modeling a galaxy's luminosity more explicitly in terms of evolving stellar populations, but they find that only the most recent star formation episode contributes substantially to the overall luminosity, validating our simple approach (see their Fig.~4).  Like our model, these studies also predict rapid declines in the galaxy populations at $z \ga 12$, making it difficult for JWST to observe significant numbers of galaxies at very high redshifts.

There are also a variety of more computationally-intensive approaches to modeling the high-$z$ galaxy population.  For example, the DRAGONS program has constructed a semi-analytic model, embedded in numerical simulations of structure formation and reionization, to describe these objects \citep{liu16,mutch16}. It includes much more of the detailed physics of galaxy formation, including mergers, infall, a multiphase ISM, and recycling. The various free parameters of the model are calibrated by matching to the observed stellar mass function at $z=5$--7. The stellar mass-halo mass relation from this model agrees reasonably well with empirical fits (see Fig.~11 of \citealt{liu16}), except at the brightest end and so also agrees with our model reasonably well.  That team has not fully explored the galaxy population at $z>10$, but they have found significantly more bright galaxies at $z \sim 11$ than our model predicts \citep{mutch16b}, in better agreement with the object from \citet{oesch16}.

Additionally, several numerical simulations have now studied high-$z$ galaxy populations (e.g., \citealt{jaacks12, trac15, feng16, gnedin16}). In general, these have subgrid star formation prescriptions calibrated to reproduce some set of observations, often at lower redshift, so their physical inputs are difficult to compare in detail with our model.  We will focus on a couple of the more informative comparisons here.  We have drawn on the dark matter measurements of \citet{trac15}, who also use abundance matching to make predictions within their simulation framework, with results generally consistent with ours. \citet{feng16} studied distant galaxies in the BlueTides simulation (notably, they also included AGN), finding that their stellar feedback prescription was the most important tunable parameter in regulating the star formation rate of faint galaxies. Their predictions at very high redshifts follow the same qualitative trends as ours, though they do not decline at high redshifts as fast as some of our models \citep{waters16}.  They also find that the bright end declines significantly less rapidly than a Schechter function, reducing the need for quenching in our model.  This implies that surveys of bright objects, as will be possible with WFIRST, will be instrumental in understanding the complex astrophysics of massive, high-$z$ galaxies. 

Finally, detailed simulations of individual galaxies with state-of-the-art feedback prescriptions produce time-averaged star formation efficiencies qualitatively similar to our feedback models. Notably, \citet{kimm14} simulated low-mass galaxies at $z \ga 6$ using a detailed prescription for internal feedback. They found that (in our language) $\tilde{f}_\star \propto m^{1/2}$, within the range of feedback models we have prescribed.  However, they also found that star formation was very bursty in low-mass systems, as the star formation ``overshot" and the subsequent supernova suppressed star formation for a substantial period.  Within our implementation, this is similar to our merger model, which we have shown also fits the data reasonably well, though of course it affects our parameter interpretation. More study is therefore needed in order to understand this star formation phase and the transition from bursty to smooth star formation.

\begin{figure}
	\includegraphics[width=\columnwidth]{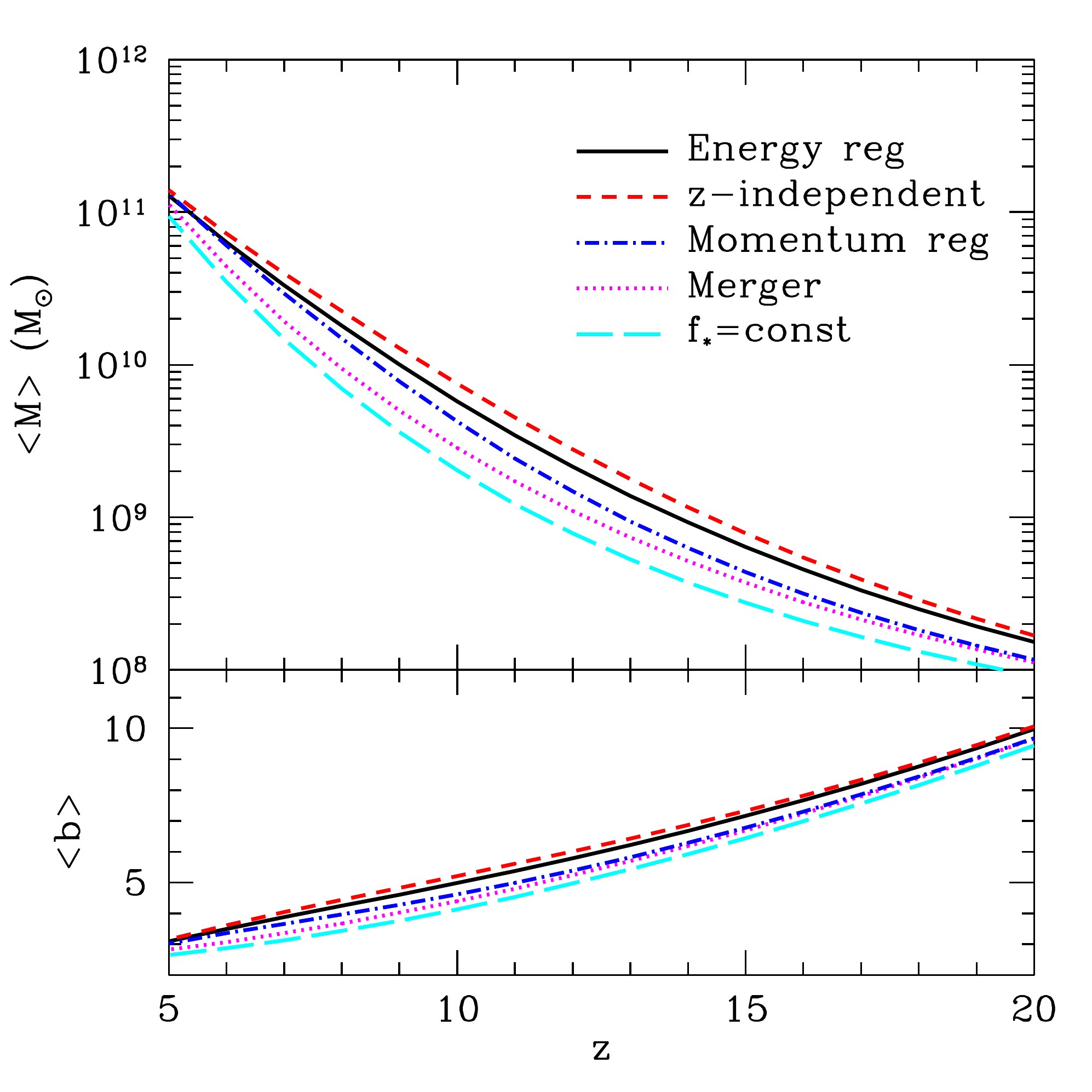}
    \caption{Luminosity-weighted average mass (upper panel) and bias (lower panel) of star-forming galaxies as a function of redshift. The solid, dotted, dashed, and dash-dotted curves show our model predictions at $z=15$ for our energy-driven feedback, momentum-driven feedback, merger-driven, and redshift-independent models. }
    \label{fig:char_q}
\end{figure}

\section{Galaxies at $z>6$} \label{gals}

We next consider some more detailed properties of galaxies at $z > 6$ according to our model.

\subsection{Average properties of star-forming halos} \label{avg}

Within our suite of models, it is useful to consider the aggregate properties of the galaxy population.  The luminosity-weighted average halo masses in our fiducial models are shown in the upper panel of Figure~\ref{fig:char_q}. The solid, short-dashed, dot-dashed, and dotted curves show results for our energy-regulated, redshift-independent, momentum-regulated, and merger models, respectively.  In all of these cases, the increasing star formation efficiency with mass (at least up to $m \approx 10^{11} \, M_\odot$ increases this population's characteristic mass by a factor of several over a case in which $f_\star$ is independent of mass (shown by the long-dashed curve), a fairly modest increase considering the wide range of halo masses available for star-forming galaxies.  This simply reflects the rapid decrease in halo number density with mass over most of the relevant parameter space.  

The lower panel shows the luminosity-weighted average bias. In all of the models, this decreases from $\sim 10$ to $\sim 3$ over this redshift interval.  In this picture of high-$z$ galaxy formation, the sources contributing most of the luminosity density are always highly clustered.  Again, the details of the feedback prescription make only a small difference to this bias.

Our different models yield similar predictions for the luminosity-weighted average halo mass (and hence bias) of the galaxy population: as we found with the star formation history, calibrating to the $z \la 8$ luminosity functions leaves relatively little freedom in the extrapolation to higher redshifts.  In this case, the average is fixed by the competition between the steeply falling halo mass function and the rising star formation efficiency (at small halo masses), so even the redshift-independent star formation efficiency model is only modestly different.  Note, however, how rapidly $\VEV{M}$ falls with redshift: it is $\la 10^9 \, M_\odot$ by $z \sim 15$, which corresponds to \emph{stellar} masses $\sim 3 \times 10^6 \, M_\odot$, illustrating the difficulty of detecting the majority of the light at extremely high redshifts.

\begin{figure*}
	\includegraphics[width=\columnwidth]{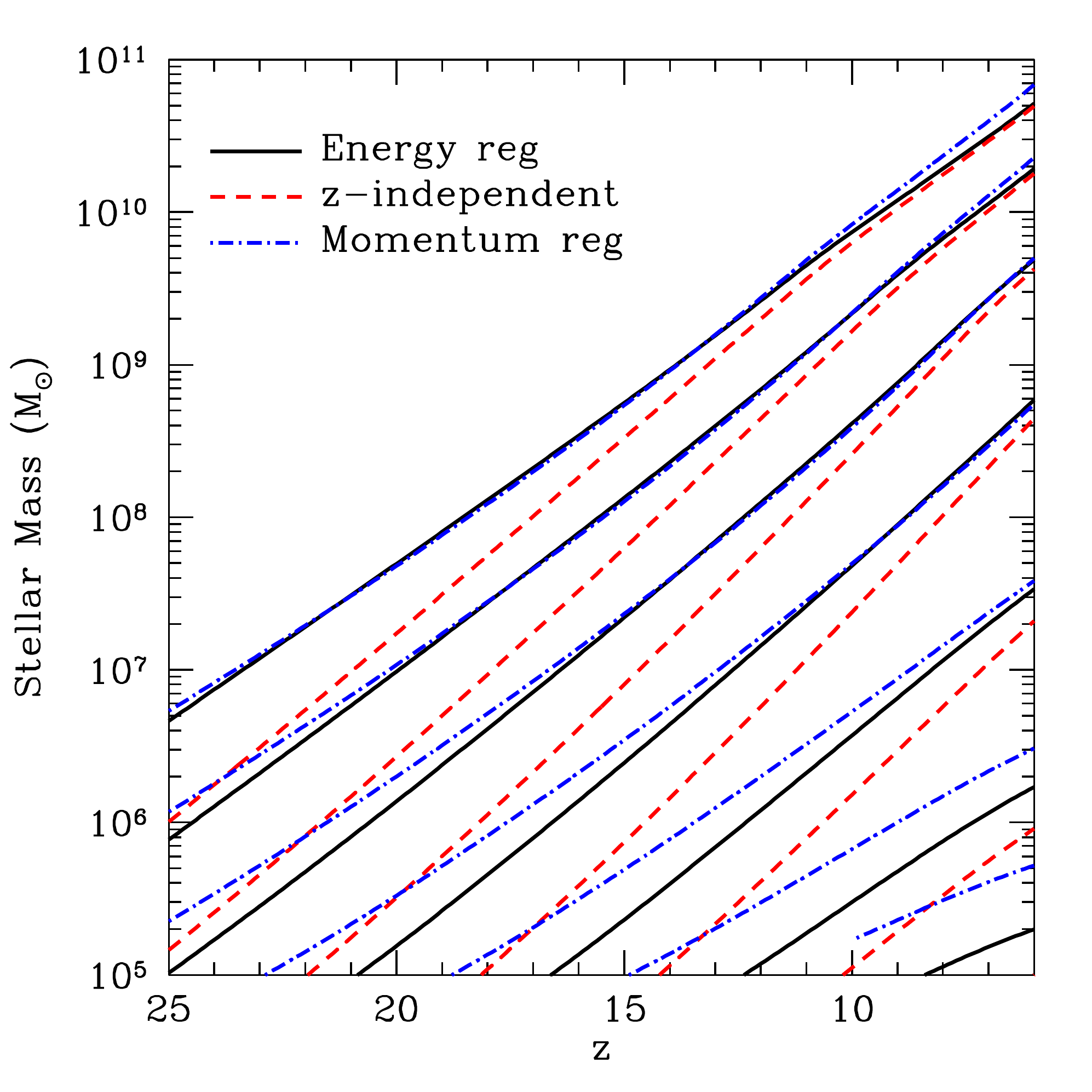} \includegraphics[width=\columnwidth]{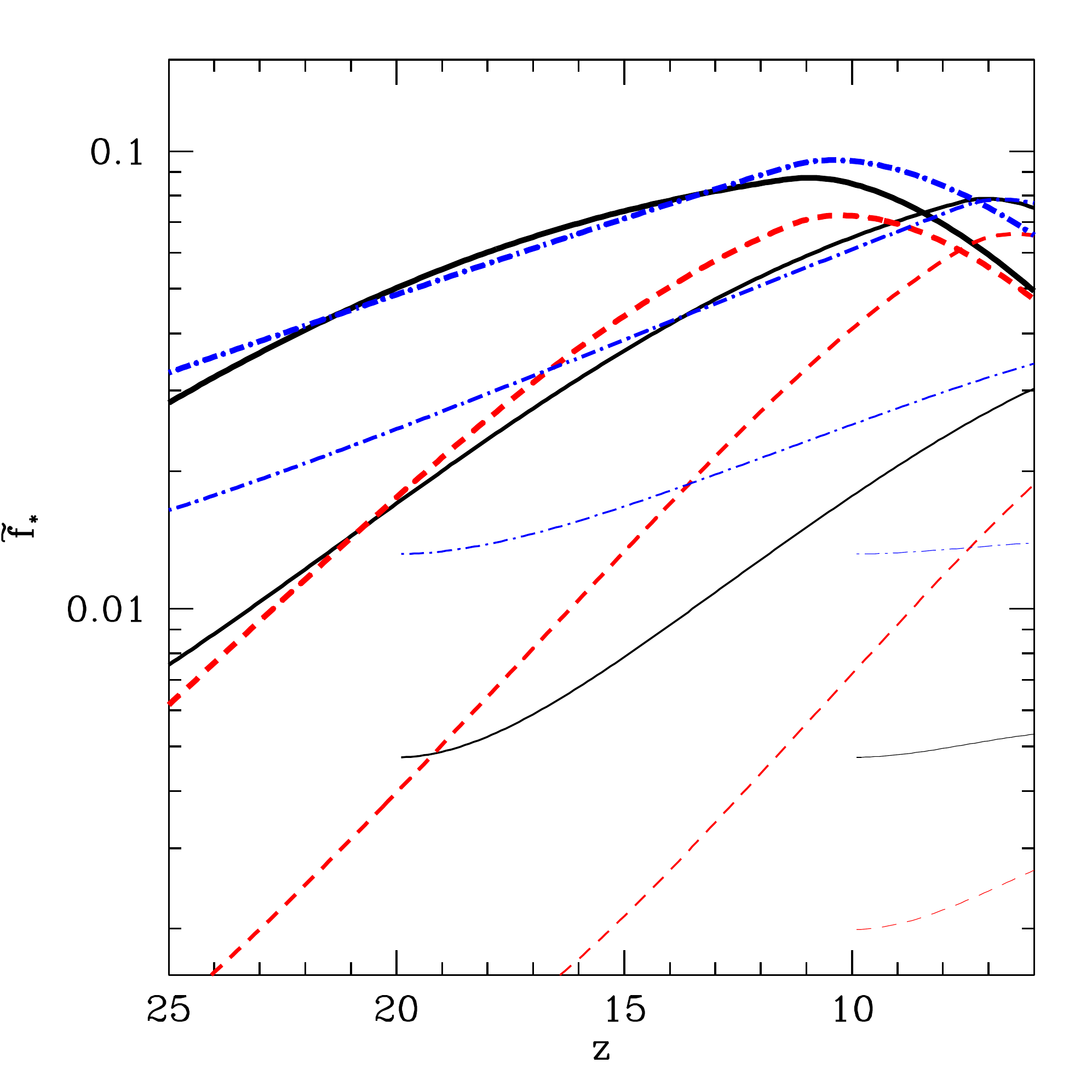}
    \caption{Star formation histories of individual halos. \emph{Left:} Stellar mass growth. The curves are initialized with masses at the atomic cooling threshold and allowed to grow according to our star formation prescriptions.  We show curves beginning at $z=40,\,35,\,30,\,25,\,20,\,15$, and 10, which have final halo masses of $M=(664,\,196,\,41,\, 6.0,\,0.71,\,0.10,\,0.024) \times 10^{10} \, M_\odot$ at $z=6$, respectively.  The solid, dot-dashed, and dashed curves show results for our energy-driven feedback, momentum-driven feedback, and redshift-independent models, respectively. \emph{Right:} Ratio of stellar mass to expected baryonic halo mass, $\tilde{f}_\star$, for a selection of halos shown at left.  For clarity, we only show halos initialized at $z=40,\, 30,\, 20$, and 10. }
    \label{fig:halo_history}
\end{figure*}

\subsection{The star formation history of individual $z>6$ galaxies} \label{history}

Because our model tracks the growth of individual halos, it allows us to follow the star formation histories of these haloes over time.  Figure~\ref{fig:halo_history} shows several examples.  We follow halos that begin forming stars at the atomic cooling threshold at a specified redshift and grow according to our abundance matching model to $z=6$.  In the left panel, we show the evolving stellar masses for halos that begin at $z=40,\,35,\,30,\,25,\,20,\,15$, and 10 and have final halo masses of $M=(664,\,196,\,41,\, 6.0,\,0.71,\,0.10,\,0.024) \times 10^{10} \, M_\odot$ at $z=6$, respectively.\footnote{We note again that these stellar masses are only approximate, as we have not accounted for the finite lifetimes of the stars.  In reality, high-mass stars would have already ended their lives, so  the remaining masses in stars will be slightly smaller.} The solid, dot-dashed, and dashed curves show results for our energy-driven feedback, momentum-driven feedback, and redshift-independent models, respectively. 

The right panel includes a subset of these models (beginning at $z=40,\, 30,\, 20$, and 10) but shows how $\tilde{f}_\star$ evolves with time.  These have final halo masses of   $M=(664,\,41,\,0.71,\,0.024) \times 10^{10} \, M_\odot$ at $z=6$, respectively.  As halos grow and their potential wells deepen, they become more resistant to feedback, and their star formation efficiencies increase by a factor of several.  In our fiducial energy-regulated and momentum-regulated models, the increasing mass is moderated somewhat by the increasing halo virial radius (at fixed mass), so the star formation efficiencies in the former two models are higher at early times, or in other words the halo stellar masses grow more slowly.  (The flatness of the curves in the right panel near their origin points is because we assume that halos immediately reach their maximal star formation rate as soon as they pass the atomic cooling threshold, so this transient behavior is an artifact of our simple model.)  Note the wide disparity in initial star formation efficiencies within the model: the weaker feedback in the momentum-regulated model allows about twice as many stars to be formed in the lowest mass halos, while the redshift-independent model falls far below the others at these very small masses.

Turning back to the left panel in Figure~\ref{fig:halo_history}, we see that rapid halo growth (see Fig.~\ref{fig:acc-rate}) and the increasing star formation rates with halo mass combine to make the stellar populations grow extremely rapidly.  All of our models are at least roughly exponentially growing, though the relevant timescale depends on the details of the model and the halo mass.  A good rule of thumb for most of these models, however, is that the stellar mass-doubling time corresponds to just $\Delta z \sim 1$--2 -- or a fraction of a Hubble time.  This is far faster than the gentle increase in the star formation efficiency and is driven by the growth of halos: early generations of galaxies have exponentially \emph{growing} star formation histories, if accretion is smooth and steady.  Even in the case of merger-driven models (or bursty feedback-driven star formation), the rapid merger rate suggests that exponentially-increasing histories are a reasonable approximation, unless the individual episodes can be resolved through high-quality observations.

This extremely rapid evolution does introduce one caveat to our model: we have assumed that feedback-regulation imposes a quasi-steady-state on star formation, but in reality that will be hard to achieve as the halos evolve.  Instead, the process of accretion onto the ISM and star formation inside clouds likely induces a delay into the response which will manifest itself as an offset from the expected $f_\star$. However, given the other uncertainties in our feedback prescriptions, we have chosen not to attempt to model this here.

\section{Implications of feedback-regulated star formation} \label{model-implications}

In this section we place our picture in the context of some of the markers of and global events induced by the first generations of galaxies in the Universe.  Following our general approach, we emphasize simple treatments of the physics to provide order-of-magnitude estimates rather than detailed models.

\subsection{The reionization history} \label{reion}

The hallmark event of this era in structure formation is the reionization of the IGM.  Even the simplest reionization models require three ingredients: (1) a description of the sources of ionizing photons, (2) a prescription for the escape of those photons into the IGM, and (3) a model for recombinations within the IGM. The first follows directly from our model, given assumptions about the stellar populations (including their metallicity, binarity, and IMF; e.g., \citealt{robertson13,mirocha16}).  We will write 
\begin{equation}
\dot{n}_{\rm ion} = N_i \VEV{f_\star} f_{\rm coll} \bar{n}_{{\rm H},0},
\label{eq:ionlum}
\end{equation}
where $\dot{n}_{\rm ion}$ is the (comoving) number density of ionizing photons produced per second,\footnote{We also include a correction factor for helium, $A_{\rm He} = 4/(4 - 3Y_p) = 1.22$, which we assume to be singly ionized along with hydrogen.} $N_i$ is the number of ionizing photons produced per baryon in stars, $\bar{n}_{{\rm H},0}$ is the comoving number density of hydrogen, and the average $\VEV{f_\star}$ is over the halo mass function.  We take $N_i = 6,000$, which correspond to a Salpeter IMF with low metallicity (see below, and Fig.~A1 of \citealt{mirocha16}). Uncertainty in the IMF and metallicity lead to at least a factor of two uncertainty in this number, though with our normalization to the observed UV luminosity function (albeit at somewhat redder wavelengths than the Lyman continuum) the effect on our models is much smaller because the star formation efficiency also depends on those assumptions \citep{mirocha16}.

\begin{figure}
	\includegraphics[width=\columnwidth]{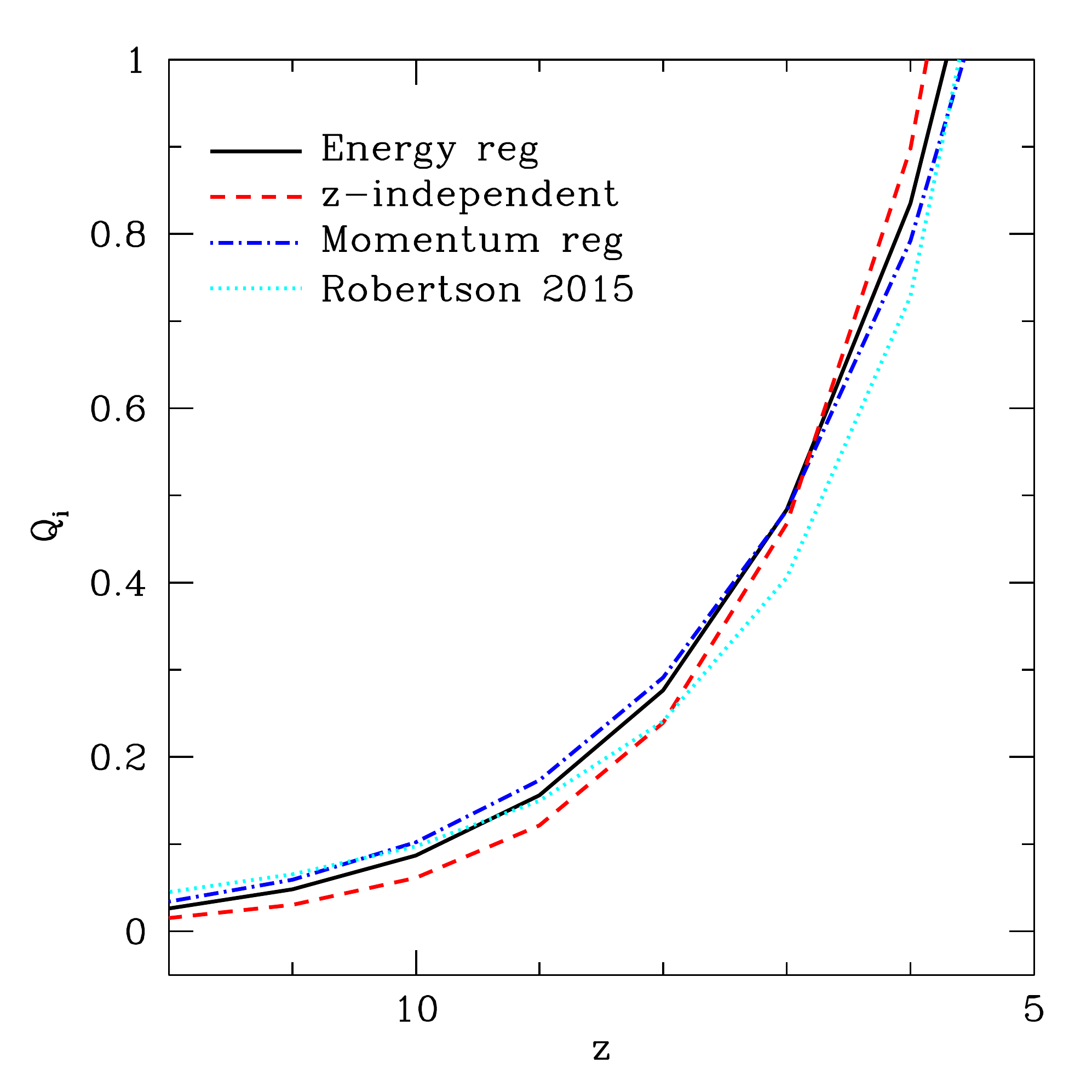}
    \caption{ Reionization histories in our models of star formation, all normalized to have $\tau_e = 0.055$. The solid, dashed, and dot-dashed curves are our energy-regulated, redshift-independent, and momentum-regulated models, while the dotted curve shows the empirical fit from \citet{robertson15}. We have varied the escape fraction in order to normalize $\tau_e$, choosing $f_{\rm esc} \approx 0.096,\, 0.084,\, 0.13,\,$ and $0.09$, respectively.}
    \label{fig:reion_history}
\end{figure}

The second ingredient is usually parameterized by the escape fraction of UV photons, $f_{\rm esc}$.  This appears to fluctuate rapidly within and between galaxies, and there is no direct observational constraint on the high-$z$ galaxy population.  In the spirit of our simple approach, we will assume that it is a constant, although there is clearly reason to believe that reality is more complex (e.g., \citealt{kuhlen12, alvarez12, sun16}).  

For the IGM, we will follow the usual clumping factor prescription, assuming that the overall recombination rate is enhanced relative to that in a uniform-density IGM by a factor $C = \VEV{n_e^2}/\VEV{n_e}^2$, where $n_e$ is the electron density and the average is taken over all ionized regions.  We will assume $C=3$ for simplicity, which is a reasonable match to simulations in the relevant redshift range \citep{pawlik09}.  We also take a fixed IGM temperature of $T = 2 \times 10^4$~K, and we use the case-B recombination rate within the ionized gas.

In that case, we can compute the ionization history via
\begin{equation}
{dQ_i \over dt} = {f_{\rm esc} \dot{n}_{\rm ion} \over \bar{n}_{{\rm H},0}} - {Q_i \over t_{\rm rec}},
\label{eq:ionhist}
\end{equation}
where $Q_i$ is the mass fraction of ionized gas and $t_{\rm rec}$ is the recombination time.  

An important constraint on the reionization history comes from the cumulative optical depth to CMB scattering:
\begin{equation}
\tau_e = c \sigma_T \bar{n}_{\rm H,0} \int_0^z f_e(z') Q_i(z') { (1+z')^2 \over H(z') } dz',
\label{eq:optdepth}
\end{equation}
where $\sigma_T$ is the Thomson cross section and $f_e(z)$ is the number of free electrons per hydrogen atom.  We assume that helium is doubly ionized at $z < 4$ and that the fraction of singly-ionized helium traces $Q_i$ at higher redshifts. Most recently, \citet{planck16-reion} measured $\tau_e = 0.055 \pm 0.009$.

We compute the reionization histories corresponding to our star formation models and show some results in Figure~\ref{fig:reion_history}. For a fair comparison, in each case we choose $f_{\rm esc}$ such that $\tau_e=0.055$, the best-fit value from Planck's measurement. While the resulting $f_{\rm esc}$ values are illustrative of the \emph{relative} production rate of ionizing photons in our models, their numerical values should not be taken too seriously, as they are degenerate with uncertainties in, for example, the stellar populations (through $N_i$ in eq.~\ref{eq:ionlum}). In Figure~\ref{fig:reion_history}, the solid, dashed, and dot-dashed curves are our energy-regulated, redshift-independent, and momentum-regulated models (which require $f_{\rm esc} \approx 0.096,\, 0.084,\, 0.13,\,$ and $0.09$ for our normalization, respectively).  For comparison, we also show the empirically-motivated model history from \citet{robertson15} with the dotted line, which requires $f_{\rm esc} \approx 0.15$ to match this $\tau_e$. 

Given the overall similarities in the star formation histories of these models, it is not surprising to see that they imply very similar reionization histories as well (at least once they are normalized in this manner). The redshift-independent model requires a higher escape fraction and has a steeper evolution: note that at $z=12$, it has roughly half the ionized fraction of the Robertson model. We conclude that -- as others have shown -- the star formation history inferred from galaxy observations is consistent with current optical depth constraints so long as a modestly higher escape fraction is assumed than is measured in lower-$z$ samples \citep{robertson13, robertson15, bouwens15-reion, sun16}. The differences amongst our models become large only at very high redshifts, when $Q_i$ is so small that it will be very difficult to measure directly.

\begin{figure}
	\includegraphics[width=\columnwidth]{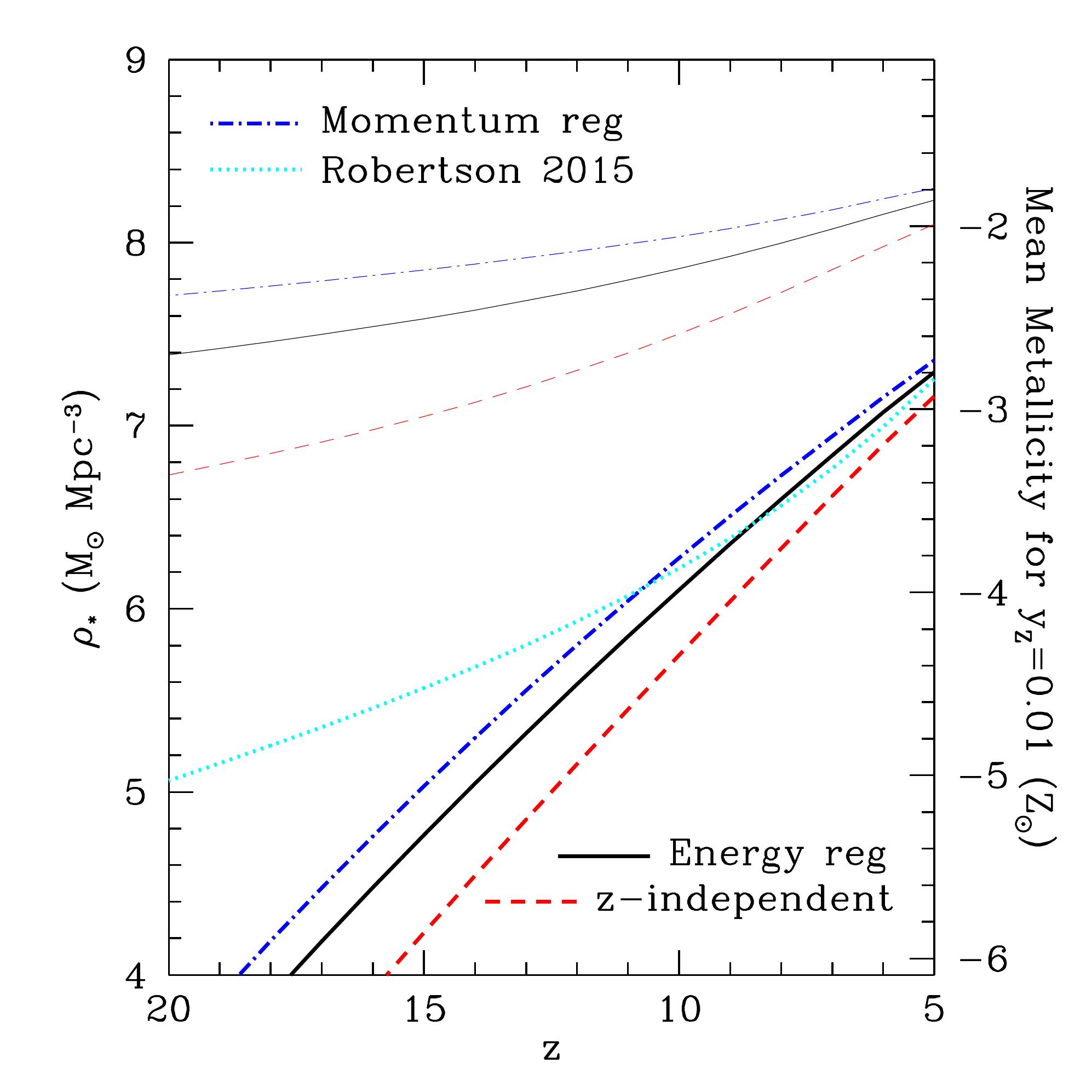} 
    \caption{ Stellar mass density in our models \emph{(left axis)} and corresponding mean metallicities \emph{(right axis}).  The latter assumes yields typical of Population II supernovae, $y =0.01$, with a standard Salpeter IMF; Population I stars would increase the overall metallicity by a factor $\la 3$, while top-heavy IMFs would increase it. The thick lines show the stellar mass density as well as the mean metallicity \emph{of the Universe}.  The thin curves show the mean metallicity of collapsed objects, \emph{if they retain all their metals}.  The solid, dashed, and dot-dashed curves are our energy-regulated, redshift-independent, and momentum-regulated models, while the dotted curve shows the empirical fit to the star formation history from \citet{robertson15}. }
    \label{fig:metal_history}
\end{figure}

\subsection{Metal production and enrichment} \label{metals}

A final, straightforward implication of the star formation history is a baseline chemical enrichment history.  The time-dependent metal yield $y_Z(t)$, or the fraction of stellar mass returned to the ISM as metals,  depends upon the IMF, stellar evolution parameters (metallicity, binarity, etc.), stellar winds, and supernova properties.  It is a function of the time delay since a stellar population formed, but after several million years $y_Z \sim 0.01$ in most cases.  With it, one can compute the rate at which metals are produced inside our galaxies.  Again, for a simple estimate we will ignore the time-dependent factor (which, after a few tens of millions of years, only depends upon winds from highly-evolved low mass stars) by including only metals from supernovae (the instantaneous recycling approximation).  In this case, the mean metallicity of the Universe can be written
\begin{equation}
\VEV{Z(z)} = {y_Z \rho_\star (z) \over \bar{\rho}_b},
\label{eq:metal-z}
\end{equation}
where $\rho_\star$ is simply the integral of the star formation rate density shown in Fig.~\ref{fig:sfhistory} and $\bar{\rho}_b$ is the mean baryon density.

Figure~\ref{fig:metal_history} illustrates the evolution of the mean metallicity in our models. To begin, the thick curves show the total stellar mass density (left vertical axis), with the solid, dashed, and dot-dashed curves corresponding to the energy-regulated, redshift-independent, and momentum-regulated models. We compare these to the empirical fit to the star formation history from \citet{robertson15} with the dotted curve.  Given their normalization to the luminosity function, it is not surprising that our results are also in agreement with stellar mass measurements in this era, though those still have large errors (e.g., \citealt{stark13}). 

On the right axis, we then convert $\rho_\star(z)$ into the mean metallicity.  Here we assume $y_Z=0.01$. This is typical of the metal yield for Population II stars with a ``normal" IMF; higher-metallicity stars will typically produce more metals overall ($y \approx 0.03$ for Population I stars) \citep{benson10}, so these values are somewhat conservative. (Moreover, a top-heavy IMF would of course produce more supernova explosions and therefore more metals.)   The thick curves show the overall metallicity of the Universe, averaged over all the baryons.  Our models have $\VEV{Z} \la 10^{-3} \, Z_\odot$ at $z \sim 6$, which is typical of any model that reionizes the Universe at that time, assuming an escape fraction $\sim 0.1$.  This is because the massive stars responsible for reionizing the IGM are (mostly) the same ones that explode as supernova and hence enrich the Universe \citep{oh_sz}, so (given present observational constraints on reionization) the overall metallicity cannot differ dramatically from this value at $z \sim 6$.

This mean value is only part of the story, however, as the real distribution is likely highly inhomogeneous. Many of these metals will likely be retained by their host galaxies: the thin curves show the average metallicity of star-forming halos, assuming that those halos \emph{retain all of their gas and metals}: this is therefore a \emph{maximal} estimate of the metallicity in these regions. To obtain the thin curves we have simply divided the thick curves by the fraction of baryons nominally inside of star-forming halos, $f_{\rm coll}$.  

Overall, we find that the mean metallicity of collapsed objects is $Z \sim 0.001$--$0.01 \, Z_\odot$, even if they retain all their metals.  This is because most of the collapsed matter is inside very small halos, where feedback strongly suppresses star formation.  Of course the actual metallicity of any given halo depends on its star formation history (and hence mass in our model): applying the analog of equation~(\ref{eq:metal-z}) to an individual galaxy, we would have $Z_{\rm gal} =  2 (y_Z/0.01) \tilde{f}_\star\, Z_\odot$.  Thus, from a glance at Fig.~\ref{fig:halo_history}, it is clear that \emph{massive} halos will have $Z \ga 0.1 \, Z_\odot$ by $z \sim 10$.

A related, but much more difficult, question is how these metals are dispersed through the IGM.  In the standard picture, galactic winds -- driven by the same feedback mechanisms we have used to regulate star formation -- eject some fraction of the metals from galaxies and advect them through the IGM. However, the efficacy of this process depends  on the wind energetics and mechanism.  Strictly following the assumptions of our energy-regulated model, the wind energy exactly balances the binding energy of the halo, and the wind material will only barely escape.  Of course, these winds are likely complex phenomena with wide velocity and density distributions within the wind material, so some of it can escape. However, even a simple estimate of the extent of these winds requires additional assumptions.

\begin{figure}
	\includegraphics[width=\columnwidth]{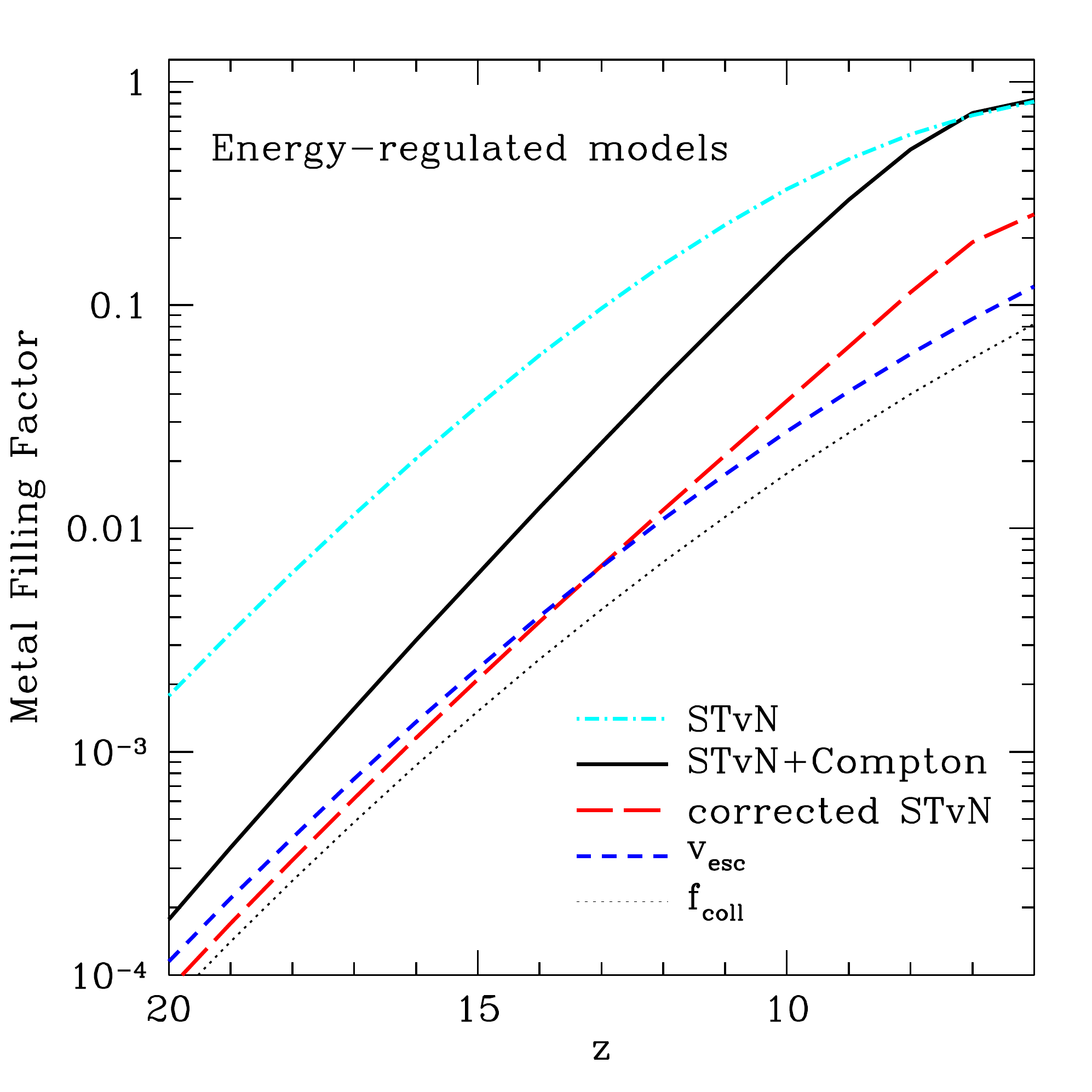} 
    \caption{ Simple estimates of metal enrichment in our star formation models. The thin dotted line shows the fraction of matter inside collapsed halos according to the \citet{trac15} mass function. The short-dashed curve assumes that metals propagate at their source halo's $v_{\rm esc}$ for half the age of the Universe. The other curves use our energy-regulated feedback model and the Sedov-Taylor-von Neumann blastwave solution. The dot-dashed line shows the maximum expansion allowed over half the age of the Universe, assuming no radiative or other losses. The solid curve includes energy loss due to Compton cooling, and the long-dashed curve includes a further approximate correction to match the results of more detailed calculations at late times. }
    \label{fig:metal_enrich}
\end{figure}

We will therefore attempt to bracket the importance of metal enrichment with some very simple estimates, which we illustrate in Figure~\ref{fig:metal_enrich}.  Each thick curve shows the fraction of the Universe's baryons that have been exposed to enriched material. To compute this fraction, we integrate the enriched volume $Q_e$ over all star-forming galaxies, according to the following wind prescriptions.  We then allow for overlap between the sources by plotting $(1-\exp^{-Q_e})$; in reality, overlap is more complicated because of source clustering \citep{furl05-double}.  For comparison, we also show the mass fraction of material that has been incorporated into star-forming halos with the thin dotted curve.\footnote{Of course, the \emph{volume} fraction of collapsed material is much smaller than this, because it is at an overdensity of $\sim 200$.} 

First, in our simplest model we suppose that the material is ejected from its source halo at that halo's escape velocity.  If it suffers no further deceleration, the comoving distance that such material would reach is
\begin{equation}
 r \approx 0.14 \ {\rm Mpc} \ \left( {7 \over 1+z} \right)^{1/2} \left( {v_{\rm esc} \over 40 \ {\rm km \, s}^{-1}} \right).
\label{eq:vballistic}
\end{equation}
We show the resulting filling fraction with the short dashed line in Figure~\ref{fig:metal_enrich} (note that this prescription does not use any of the details of our star-formation prescriptions). These kinds of slow winds near the escape velocity result in only very modest enrichment. 

However, the energy available from feedback can lead to much faster expansion, if it has only a few times more energy than required in our model.  The dot-dashed curve in Figure~\ref{fig:metal_enrich} is the most extreme case we consider. Here, we assume that all the energy generated by feedback is used to drive a blastwave into the surrounding IGM and neglect the gravitational potential of the host halo, so that the wind follows the energy-conserving Sedov-Taylor-von Neumann solution. In this case, $r \propto (E t^2 / \bar{\rho}$.  For simplicity we assume the wind propagates into a medium at the mean density of the Universe, $\bar{\rho}(z)$, in which the wind has been active for half the age of the Universe. In this case, the comoving wind radius in our energy-regulated model is \citep{voit96}
\begin{equation}
r \approx 0.25 \left( { m_h \over 10^8 \ M_\odot} \right)^{1/3},
\label{eq:rstvn}
\end{equation}
Interestingly, in this case the enriched mass is independent of the source's redshift and proportional to the source's mass, with each halo enriching $\sim 25$ times its own mass.\footnote{Note that a very similar estimate results from assuming that the blastwaves expand until they have accelerated all their material to the local Hubble flow velocity \citep{furl01}.}

This is certainly an overestimate, however, because the blastwave will undergo radiative (and other) losses as it propagates.  Detailed wind models are beyond the scope of this paper (see, e.g., \citealt{ostriker88, tegmark93, furl03-metals}), but we provide two simple corrections.  First, the blastwaves will inevitably lose energy to Compton cooling off the CMB, for which the cooling time is $t_{\rm Comp} \approx 1.2 \times 10^8 [10/(1+z)]^4$~yr.  We approximate this effect by limiting winds to expand for no more than $t_{\rm Comp}$ and show the result with the solid curve in Figure~\ref{fig:metal_enrich}.  This sharply reduces the enriched volume at high redshifts, when Compton cooling is efficient.  Finally, detailed wind models show that our simple treatment typically overestimates the maximum radius by a factor of $\sim 2$ due to other losses \citep{furl03-metals}.  The long-dashed curve in Figure~\ref{fig:metal_enrich} shows this corrected filling factor. Note that the fraction does not always fall by a factor of 8 because we always assume that material inside star-forming halos is enriched.

We see that the enrichment process is likely very inhomogeneous.  Only at $z \la 10$ can a substantial fraction of the volume be enriched, and in our ``best guess" models the fraction is still $\la 20\%$.

\section{Discussion} \label{disc}

We have presented a simple framework for modeling the high-redshift galaxy population.  The three ingredients are the dark matter halo abundance, their average accretion rates, and a prescription for setting the star formation efficiency $f_\star$ through stellar feedback.  We roughly calibrate the free parameters in the feedback scheme by comparison to the measured galaxy luminosity functions at $z=6$--10, finding that a variety of feedback prescriptions fit the currently available data with reasonable parameter choices.  In all cases, we find that $f_\star$ peaks at $m_h \sim 10^{11}$--$10^{12} \, M_\odot$, possibly declining toward larger masses and certainly declining rapidly at lower masses.  This is consistent with empirical fits using similar schemes (e.g., \citealt{mason15, mashian16, sun16, mirocha16}).  Overall we find that halos turn $\sim 1$--10\% of their total baryonic mass into stars, with a strong dependence on halo mass.

With our model, we then extrapolate to higher redshifts and fainter galaxies with a clear understanding of the physical meaning of the extrapolation. Necessarily, our extrapolation implicitly assumes that the underlying mechanisms of galaxy and star formation do not evolve with redshift, which is certainly too simplistic.  But it provides a baseline against which evidence for new physics can be evaluated.  Interestingly, the data are already good enough to estimate the amount of star formation beyond $z \sim 10$, even allowing for variations in our model's parameters (provided again that the underlying physics does not change).  Within our parameterization of feedback, the most significant question is whether the processes controlling star formation depend only on halo mass or if we allow explicit dependence on redshift.  The $z < 10$ data are consistent with a redshift-independent solution, and in fact the ``natural" redshift dependence of feedback models (in which the star formation efficiency of halos at a fixed mass increases with redshift, as the halos are more tightly bound at early times) appears to overpredict modestly the number of $z \sim 9$--10 galaxies. 

However, \citet{oesch16} recently discovered an extremely bright source at $z=11.1$.  The implied number density of similar objects is very large (albeit with significant errors for a single detection), and even the most optimistic of our models -- which assumes strong redshift dependence within the feedback -- struggles to reproduce this object, which requires very strong star formation in massive halos.  However, it is worth noting that this object tells us little about the dominant star formation mode, as it is such a massive galaxy that it is likely unaffected by stellar feedback (which we assume controls the star formation rate in the much more abundant smaller halos).  

We have shown (see Fig.~\ref{fig:lf_highz}) that future space missions are ideally suited to measure precisely the high-$z$ luminosity function of galaxies over a factor of nearly 10$^4$ in luminosity, at least at $z \la 15$.  While deep surveys with JWST will uncover the bulk of the galaxy emission, WFIRST will be essential for exploring the growth of the most massive galaxies.  However, in our model the galaxy luminosity function decreases rapidly enough that by $z \sim 15$, very few sources will be detected even in very deep surveys.  Detections in this era and beyond will therefore indicate the presence of new stellar populations or formation mechanisms.  

We have further explored some of the basic implications of our model.  Within feedback-regulated models the star formation efficiency increases with halo mass, so most of the star formation occurs in halos with virial temperatures well above the atomic cooling threshold, especially at $z \la 10$.  This implies that the smallest halos are less important than previously assumed and that ``photoheating feedback" during reionization is less important as well. Even though the star formation processes evolve only slowly (if at all) with redshift, the characteristic halo mass still evolves quickly, because massive halos are forming so rapidly during this era.   We have also found that our histories provide reasonable reionization histories, consistent with available data, if we set $f_{\rm esc} \sim 0.1$, and that they likely enrich only a small fraction of space through galactic winds.

Our model is clearly far too simplistic to provide an accurate understanding of the earliest generations of galaxies.  We ignore, for example, evolution in the IMF, chemistry, and the spatial distribution of star formation.  However, we have used its simple physical principles to model the available data and shown how it can easily be extrapolated to higher redshifts while understanding the detailed physical implications of this extrapolation.  We hope that this model will be useful in rapid explorations of the parameter space allowed to high-$z$ galaxies, in forecasting future surveys, and in qualitatively understanding the nature of these sources.  In the future, we will explore further implications of our model for distant galaxies as well as improve its physical inputs.

\section*{Acknowledgements}

SRF thanks A. Benson, A. Dekel, and B. Robertson for helpful conversations. This work was supported by the National Science Foundation through award AST-1440343 and by NASA through award NNX15AK80G. We also acknowledge a NASA contract supporting the ``WFIRST Extragalactic Potential Observations (EXPO) Science Investigation Team" (15-WFIRST15-0004), administered by GSFC. SRF was  partially supported by a Simons Fellowship in Theoretical Physics and thanks the Observatories of the Carnegie Institute of Washington for hospitality while much of this work was completed. 







\appendix

\section{The formation of new galaxies} \label{bdryterm}

Ongoing accretion onto existing galaxies is not the only channel through which stars can form at very high redshifts: in particular, if one imposes a mass threshold $m_{\rm min}$, the steepness of the mass function implies that halos passing that threshold can carry a non-negligible fraction of the overall increase in galaxy mass density. To see this, note that in our picture the total rate at which the collapse fraction $f_{\rm coll}$ (or the fraction of mass inside collapsed objects) changes is the sum of two terms:
\begin{equation}
\bar{\rho} {d f_{\rm coll} \over dt} = \int_{m_{\rm min}}^\infty dm \, \left( {\dot{m} \over \bar{\rho}} \right) n_h(m,z) + \left[ \dot{m}  (mn) \right]_{m_{\rm min}},
\label{eq:fcoll-deriv}
\end{equation}
where the first term on the right-hand side describes accretion onto halos already above the threshold and the second term describes halos that just cross the threshold.\footnote{For clarity of presentation, in the remainder of this section we will assume that $m_{\rm min}$ is constant with redshift, although in practice it varies slowly with time.} This expression follows from taking the total time derivative of the integral and noting that the total number of halos is conserved in our ``abundance-matching" picture, with only the mass of each individual halo evolving over time.

\begin{figure}
	\includegraphics[width=\columnwidth]{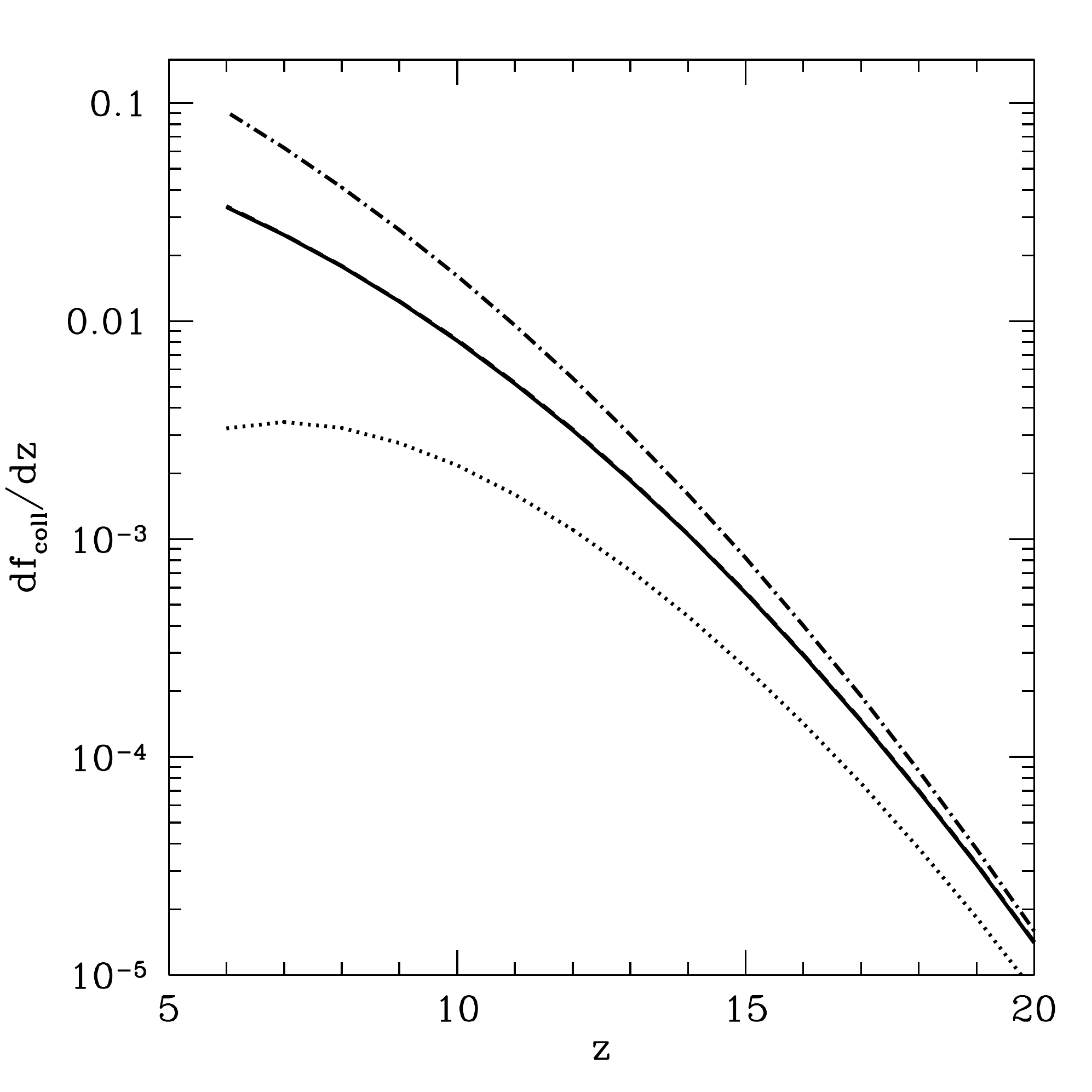}
    \caption{The importance of halos crossing the minimum mass threshold. The solid line shows $d f_{\rm coll}/dz$, the dashed line (nearly indistinguishable from the solid line) shows the sum of the two terms on the right-hand side of equation~(\ref{eq:fcoll-deriv}), and the dotted line shows the boundary term alone. For reference, the dot-dashed line shows $f_{\rm coll}(z)$ in our model. All curves assume that $m_{\rm min} = 10^8 \ M_\odot$.}
    \label{fig:acc-comp}
\end{figure}

Figure~\ref{fig:acc-comp} shows the importance of these two terms: the solid line shows $d f_{\rm coll}/dz$, the dashed line -- which is almost indistinguishable from the solid line -- shows the sum of the two terms on the right-hand side of equation~(\ref{eq:fcoll-deriv}),\footnote{Note that our abundance matching prescription is not guaranteed to reproduce the overall evolution, because it relates only integrated quantities. However, this comparison shows that it is nevertheless very accurate.} and the dotted line shows the boundary term alone. This is a small fraction of the overall evolution at $z \sim 6$, but by $z \sim 10$ it is non-negligible, and at higher redshifts it provides about half the new mass.  For reference, the dot-dashed line shows $f_{\rm coll}(z)$ in our model, assuming $m_{\rm min} = 10^8 \ M_\odot$.  

In the main text, we ignore the boundary term when computing star formation rates, because it is not clear that our approximations or physical picture make sense in these newly-formed halos.  For example, many will have already undergone bursts of star formation seeded by Population III stars, so their initial stellar populations and gas contents will have already been disturbed.  These objects are very unlikely to substantially affect our luminosity function predictions, because the halos are so fragile to feedback.  But they will affect the global star formation rate in some way, especially at $z \ga 15$ where they carry roughly half of the overall increase in collapsed mass.

\bsp	
\label{lastpage}
\end{document}